\documentclass[a4paper,11pt]{article}
\pdfoutput=1 % if your are submitting a pdflatex (i.e. if you have
\usepackage{jheppub} % for details on the use of the package, please
\usepackage{graphicx}
\usepackage[figurename=Fig.]{caption}
\usepackage[tablename=Tab.]{caption}
\usepackage{color}
\usepackage{float}
\usepackage{amssymb}
\usepackage{amsmath}
\usepackage{mathrsfs}
\usepackage{textcomp}
\usepackage{bbm}
\usepackage{cases}
\usepackage{makecell}
\usepackage{pdflscape}
\usepackage{subfigure}
\usepackage{pdfpages}
\usepackage{makeidx}
\usepackage{bm}
\usepackage{xstring}
\usepackage{lipsum}
\usepackage{slashed}
\usepackage{multicol}
\usepackage{setspace}
\usepackage{booktabs}
\usepackage[table]{xcolor}
\usepackage{multirow}
\usepackage{braket}
\usepackage[separate-uncertainty]{siunitx}
\usepackage[displaymath, mathlines]{lineno}

\newcommand{\eref}[1]{Eq.\,(\ref{#1})}
\newcommand{\w}{\omega}
\newcommand{\vo}{\vec{o}\@ifnextchar{^}{\,}{}}
\def\up{\mathrm}
\def\d{\up{d}}

\renewcommand{\thetable}{\Roman{table}}

\graphicspath{{PDF/}}

\newcommand{\vabs}[1]{|\vec #1\,|}

\def\slash#1{\setbox0=\hbox{$#1$}           % set a box for #1 
   \dimen0=\wd0                                 % and get its size 
   \setbox1=\hbox{/} \dimen1=\wd1               % get size of / 
   \ifdim\dimen0>\dimen1                        % #1 is bigger 
      \rlap{\hbox to \dimen0{\hfil/\hfil}}      % so center / in box 
      #1                                        % and print #1 
   \else                                        % / is bigger 
      \rlap{\hbox to \dimen1{\hfil$#1$\hfil}}   % so center #1 
      /                                         % and print / 
   \fi}                                         %     
\def\sl#1{\setbox0=\hbox{#1} 
  \dimen0=\wd0 
  \rlap{\hbox to \dimen0{\hss/\hss}}% 
  #1} 
\title{\boldmath Mass spectra, wave functions and mixing effects of the $(bcq)$ baryons}

\author[1]{Qiang Li}
\author[2,3,4]{Chao-Hsi Chang}
\author[5]{Si-Xue Qin}
\author[6]{Guo-Li Wang}

\affiliation[1]{School of Physical Science and Technology, Northwestern Polytechnical University, Xi'an 710072, China}
\affiliation[2]{Institute of Theoretical Physics, Chinese Academy of Sciences, Beijing 100190, China}
\affiliation[3]{School of Physical Sciences, University of Chinese Academy of Sciences, 19A Yuquan Road, Beijing 100049, China}
\affiliation[4]{CCAST (World Laboratory), P.O. Box 8730, Beijing 100190, China}
\affiliation[5]{Department of Physics, Chongqing University, Chongqing 401331, China}
\affiliation[6]{Department of Physics, Hebei University, Baoding 071002, China}

\emailAdd{liruo@nwpu.edu.cn}
\emailAdd{zhangzx@itp.ac.cn}
\emailAdd{sqin@cqu.edu.cn}
\emailAdd{gl\_wang@hit.edu.cn}

\abstract{
Mass spectra and wave functions of the $J^P=\frac{1}{2}^+$ $(bcq)$ baryons are calculated by the relativistic Bethe-Salpeter equation\,(BSE) with considering the mixing effects between the $1^+$ and $0^+$ $(bc)$-diquarks inside. Based on the diquark picture, the three-body problem of baryons is transformed into two two-body problems. The BSE and wave functions of the $0^+$ diquark are given, and then solved numerically to obtain the effective mass spectra and form factors. Also we present the wave functions at zero point for the $(bc)$-diquark. Considering the obtained diquark form factors, the $(bcq)$ baryons are then described by the BSE as the bound state of a diquark and a light quark,  where the interaction kernel includes the inner transitions between the $0^+$ and $1^+$ diquarks. The general wave function of the $\frac{1}{2}^+$ $(bcq)$ baryons is constructed and solved to obtain the corresponding mass spectra. Especially, by using the obtained wave functions, the mixing effects between $\Xi_{bc}(\Omega_{bc})$ and $\Xi_{bc}'(\Omega'_{bc})$ in ground states are computed and determined to be small\,($\sim \!1\%$). The numerical results indicate that it is a good choice to take $\Xi_{bc}$ and $\Xi'_{bc}$ as the baryon states with the inside $(bc)$-diquarks occupying the definite spin.
}

\begin{document} 
\maketitle
\flushbottom

\section{Introduction}
In 2019, the LHCb Collaboration first reported the doubly charmed baryon $\Xi_{cc}^{++}$  with the determined mass $M_{\Xi_{cc}^{++}}=3.621$ GeV\,\cite{LHCb2017B,LHCb2020-Xicc} and lifetime $\tau_{\Xi^{++}_{cc}}=0.256\,\si{ps}$\,\cite{LHCb2018}. This is the first time to confirm a baryon consists of two charm quarks and a $u$ quark. This discovery has caused great attention in hadron physics, and led to a new round of researches on the heavy baryons, which may hint that more and more baryons with two even three heavy quarks would be detected in the near future. However,  there are still no significant signals for the isospin partner of $\Xi_{cc}^{++}$\,\cite{LHCb2020-Xicc+} and the $\Xi_{bc}$\,\cite{LHCb2020-Xibc}. $\Xi^{++}_{cc}$ contains two charm quarks which constraint the $(cc)$-diquark could only be flavor symmetric, while in the other case, the two different heavy-flavor quarks in the baryons could be both in the flavor symmetric and antisymmetric.  In this work, we would focus on the latter ones, the baryons with flavor constituents $(bcq)$ with $q$ denoting a light quark $u$, $d$, or $s$.

Different from a meson, a baryon consists of three quarks and has the $\frac{1}{2}$-integer spin  and hence obeys the Fermi-Dirac statistics principle. The Pauli principle states that the wave function of a baryon should be antisymmetric under the interchange of any two quarks inside. The baryon wave functions includes four parts, spatial, spin, color, and flavor, and it is the whole $\psi(\up{baryon})=\psi(\up{space})\psi(\up{spin})\psi(\up{color})\psi(\up{flavor})$ works that should be anti-symmetric under the interchange of any two quarks. Notice this principle also works for a diquark system inside the baryon, and here we regard all these four parts as the identical quark's different properties. Since the naturally occurring particle is always a color singlet, any two quarks inside a baryon are in the color anti-triplet in order to form a colorless particle. Hence the color part is always anti-symmetric and the others together are symmetric. In the orbital ground state, there is no angular dependence and thus the spatial part is symmetric. For a doubly heavy diquark system consisting of two quarks of the same flavor, such as the case in $\Xi_{cc}$ or $\Xi_{bb}$, there is no other choices, it is flavor symmetric and hence symmetric in spin, which means it could only be in $1^+$ state. We have discussed this case in a previous work\,\cite{LiQ2020}. However, for a doubly heavy diquark system consisting of two different quarks, namely the $(bc)$ case, its flavor could be both symmetric or anti-symmetric, namely, it could be in both $1^+$ and $0^+$ state. The former one has also been incorporated in Ref.\,\cite{LiQ2020}, and in this work first we would focus on the latter one, the doubly heavy baryons with $0^+$ $(bc)$-diquark cores; then we deal with the real physical baryons states by considering the mixing effects between the $1^+$ and $0^+$ $(bc)$-diquark cores.

The $0^+$ $(bc)$-diquark core combined with a third light quark can form the $J^P=\frac{1}{2}^+$ baryon state in $S$-wave. We would first construct the Salpeter wave functions, and then give the mass spectra and numerical wave functions by solving the corresponding three-dimensional (Bethe-)Salpeter equations under instantaneous approximation. For the $\frac{1}{2}^+$ $(bcq)$ baryon states, it can be formed by both the $0^+$ and $1^+$ $(bc)$-diquark cores, and hence usually the doublets $\Xi_{bc}(\Omega_{bc})$ and $\Xi_{bc}'(\Omega'_{bc})$ are the mixing states of these two. These mixing effects would be considered by the coupled BSE.  The masses of the $\Xi^{(\prime)}_{bc}$ and $\Omega^{(\prime)}_{bc}$ are predicted to lie in $6.8\sim 7.1\,\si{GeV}$ and $6.9\sim7.1\,\si{GeV}$ respectively \cite{SilvestreBrac1996,Ebert1997,Ebert2002,Kiselev2002B,HeDH2004,Albertus2007,Martynenko2007,Roberts2008,ZhangJR2008,Giannuzzi2009,TangL2012,Ghalenovi2014,Brown2014,Roncaglia1995,Karliner2014,Karliner2018,WengXZ2018}.
Notice that in Refs.\,\cite{Roncaglia1995,Karliner2014,Karliner2018,WengXZ2018} the $\frac{1}{2}^+$ $(bcq)$ baryon doublets are labeled by the definite spin of the $(cq)$-diquark,  while in Refs.\,\cite{Ebert1997,Ebert2002,HeDH2004,Albertus2007,Martynenko2007,Roberts2008,ZhangJR2008,Giannuzzi2009,TangL2012,Brown2014} the baryons are labeled by the $(bc)$-diquark spin.
The naming conventions used in this work are: unprimed symbols $\Xi_{bc}$ and $\Omega_{bc}$ to denote the baryon states with the dominate $1^+$ $(bc)$-diquark, while the primed symbols $\Xi'_{bc}$ and $\Omega'_{bc}$ to label the ones with the dominate $0^+$ $(bc)$-diquark. 

In this work, we will deal with the doubly heavy baryons with $(bcq)$ flavors within the framework of the instantaneous Bethe-Salpeter equation, which has already been successfully used to cope with the doubly heavy baryons with $1^+$ diquark cores\,\cite{LiQ2020},  the recently observed pentaquark $P_c$ and fully heavy tetraquark $ T_{QQ\bar Q\bar Q}$ states\,\cite{XuH2020,LiQ2021}, and also generally applied to the meson mass spectra\,\cite{Chang2005A,Chang2010,LiQ2019A}, the hadronic transitions and decays\cite{Chang2005,WangZ2012A,WangT2013,WangT2013A,LiQ2016,LiQ2017,LiQ2017A}. The theoretical calculations from BS methods have achieved satisfactory consistences with the experimental measurements, which give us more confidence and thereby inspire us  to apply the same methods in baryon bound problems.
Our basic scheme is to reduce the three-body baryon bound problem into two two-body bound problems. Namely, the two heavy quarks first form a compact diquark core in color anti-triplet, and then the diquark combined with a third quark forms the colorless baryon. In Refs.\,\cite{Cahill1987,Maris2002,Maris2004,Maris2005} the Bethe-Salpeter methods have already been used to study the properties of the diquark, although different interaction kernels are adopted. By supposing a quark exchange between diquark and the third quark, Refs.\,\cite{Keiner1996,Keiner1996A,Keiner1997} built the effective instantaneous BS equation for a baryon; while in Refs.\,\cite{GuoXH1996,GuoXH1999,GuoXH2007,WengMH2011,ZhangL2013,LiuY2015,WeiKW2017,LiuLL2017} a different scheme to deal with the baryon is adopted based on the diquark-quark picture in the heavy quark limit.

This paper is organized as follow. In Section\,\ref{Sec-2} the $0^+$ diquark mass spectra and wave function are obtained by solving the corresponding diquark BSE, and then the relevant form factors describing the diquark-gluon interaction are calculated. In Section\,\ref{Sec-3} we derive the BSE of a $\frac{1}{2}^+$ baryon as the bound state of a $0^+$ diquark and a third quark. In Section\,\ref{Sec-4} the BSE and the corresponding wave function of the general $\frac{1}{2}^+$ $(bcq)$ baryons is given including the mixing effects between the  $1^+$ and $0^+$ $(bc)$-diquark transition. In Section\,\ref{Sec-5} the mass spectra, numerical wave functions, and the mixing effects are presented for the $(bcq)$ baryons with $J^P=\frac{1}{2}^+$. Finally, we give a brief summary in Section\,\ref{Sec-6}.

\section{Bethe-Salpeter equation of the $0^+$ diquarks and the relevant form factors}\label{Sec-2}

In order to have a self-contained discussion, we first review the BS methods for a diquark and the interaction kernel used; then the $0^+$ $(bc)$-diquark Salpeter wave function is solved to calculate the relevant form factors.

\subsection{\textup{Bethe-Salpeter equation and wave function of the $0^+$ diquark core} }

%\begin{figure}[h!]
%\vspace{0.5em}
%\centering
%\includegraphics[width=0.55\textwidth]{BS-D2}
%\caption{BSE of a diquark. $p$ denotes the bound state momentum, and $p^2=\mu^2$, where $\mu$ stands for the bound system mass; $s_{1(2)}$ and $u_{1(2)}$ are the corresponding quark momenta.}\label{Fig-BSD}
%\vspace{0.5em}
%\end{figure}
In the momentum space, the diquark Bethe-Salpeter equation can be expressed as\,\cite{LiQ2020}
\begin{equation}
\Gamma_c(P_\up{D},s)=\int \frac{\d 4 u}{(2\pi)^4} iK_c(s-u) \left[ S(u_1)\Gamma_c(P_\up{D},u)S(-u_2)  \right], \label{E-BS-vertex}
\end{equation}
where $\Gamma_c$ is related to the diquark vertex $\Gamma_\up{D}$ by $\Gamma_\up{D}=\Gamma_c C$, and $C\equiv i\gamma^0\gamma^2$ denotes the charge conjugate operator; the interaction kernel $K_c(s-u)=\frac{1}{2}K_\up{M}(s-u)$, where $K_\up{M}$ denotes the interaction kernel in a meson and will be introduced more detailed later; $P_\up{D}$ is the total momentum of the diquark; $S(s_{1})$ and $S(-s_2)$ are the Dirac propagators, and $s_{1(2)}$ is the corresponding momentum. The internal momentum $s$ and $u$ are defined as
\[s=\lambda_2s_1-\lambda_1s_2,~~~u=\lambda_2 u_1-\lambda_1 u_2,\]
where $\lambda_i \equiv \frac{m_i}{m_1+m_2}~(i=1,2),$ and $m_i$ denotes the constituent quark mass. From the BS vertex, the BS wave function can be naturally defined as 
\begin{equation} \label{E-BS-D}
\psi_c(P_\up{D},s)=S(s_1)\Gamma_c(P_\up{D},s)S(-s_2).
\end{equation}
Notice the original diquark Bethe-Salpeter wave function is related to the above $\psi_c$ by $\psi_\up{D}=\psi_c C$. By solving the \eref{E-BS-vertex} we can obtain the diquark vertex $\Gamma_\up{D}$ or the diquark BS wave function $\psi_\up{D}$.
%The original diquark Bethe-Salpeter wave function behaves as
%\begin{equation} \label{E-BS-D}
%\psi_\up{D}(P_\up{D},s)=S(s_1)\Gamma_c(P_\up{D},s)S^\up{T}(s_2),
%\end{equation}
%which is related to the $\psi_c$ by $\psi_\up{D}=\psi_c C$.

Throughout this work we would study under the instantaneous approximation (IA), which states the interaction potential we used are static and does not depend on the time component of the exchanged momentum, namely, $K_c(s-u)\sim K_c(s_{\!\perp}-u_{\!\perp})$. Under this approximation, the integration over the time component of $s$ can be further absorbed into the wave function. Namely, we can define the three-dimensional Salpeter wave function as 
\begin{gather}
\varphi_c(P_\up{D},s_{\!\perp}) \equiv i\int \frac{\d s_{0}}{2\pi}\psi_c(P_\up{D},s),
\end{gather} 
where $s_{0}\equiv  \frac{s\cdot P_\up{D}}{M_\up{D}}$ and $M_\up{D}$ is the constituent mass of the diquark; $s_{\!\perp} \equiv s-s_{0} \frac{s\cdot P_\up{D}}{M_\up{D}}$. By performing the contour integral over $s_{0}$ on both sides of \eref{E-BS-D}, we  obtain the three-dimensional (Bethe-)Salpeter equation\,\cite{LiQ2020},
\begin{equation}\label{E-BS-HM}
M_\up{D} \varphi_c(P_\up{D},s_{\!\perp})= (\w_1+\w_2)  H(s_{1\perp})\varphi_c(s_{\!\perp}) +\frac{1}{2} \left[ H(s_{1\perp}) W(s_{\!\perp}) - W(s_{\!\perp})  H(s_{2\perp})\right],
\end{equation}
where the kinetic energy of the constituent quark is defined as $\w_i\equiv ({m_i^2-s_{i\perp}^2})^{\frac12}$ with $s_{i\perp}=s_i- \frac{s\cdot P_\up{D} P_\up{D}}{M^2_\up{D}}$. The dimensionless operator ${H}(s_{i\perp})$ is the usual Dirac Hamiltonian divided by $\w_i$, namely,
\begin{gather}
H(s_{i\perp}) \equiv  \frac{1}{\w_i}(\slash s_{i\perp}+m_i)\gamma^0;
\end{gather}
Also we define $W(s_{\!\perp}) \equiv \gamma^0\Gamma_c(P_\up{D},s_{\!\perp})\gamma_0$ to denote the potential energy part of the inner constituents. Notice if the Lorentz structure in the kernel $K(s_{\!\perp}-u_{\!\perp})$ is the type $\gamma^0\otimes \gamma_0$, $W(s_{\!\perp})$ would be a scalar integration, which is just the case we deal with in this work. The three-dimensional vertex is expressed as
\begin{equation}\label{E-vertex-3D}
\Gamma_c( s_{\!\perp}) =\int \frac{\d^3 u_{\!\perp}}{(2\pi)^3}  K_c(s_{\!\perp}-u_{\!\perp}) \varphi_c( u_{\!\perp}).
\end{equation}
Namely, the integration over $u_0\equiv \frac{u \cdot P_\up{D}}{M_\up{D}}$ has been absorbed into the definition of the Salpeter wave function $\varphi_c(u_{\!\perp})$, and now we only need to deal with the problem of the three-dimensional integration.

Also the Salpeter wave function $\varphi_c(P_\up{D},s_{\!\perp})$ fulfills the following constraint condition,
\begin{equation} \label{E-BS-HM2}
\hat  H(s_{1\perp}) \varphi_c(s_{\!\perp}) + \varphi_c(s_{\!\perp})\hat  H(s_{2\perp}) =0.
\end{equation}
The normalization condition of the Salpeter wave function reads
\begin{equation}
%\int \frac{\up d^3 \vec q}{(2\pi)^3}\frac{1}{2} \up{Tr}\left[ \varphi^\dagger  \hat H_1\varphi -  \varphi \hat H_2\varphi^\dagger \right]=
\int \frac{\up d^3  s_{\!\perp}}{(2\pi)^3} \, \up{tr}~ \varphi_c^{\dagger}(s_{\!\perp})  \hat H(s_{1\perp})\varphi_c(s_{\!\perp})=2M_\up{D}.
\end{equation}
Note that the results above are also available for a meson system.

Since a diquark consists of two quarks rather than the quark-antiquark pair in a meson, the diquark parity is  then just opposite with the corresponding meson. Considering the Lorentz structure, total angular momentum and parity, the three-dimensional BS wave function of a diquark with $J^P=0^+$ can be constructed as 
\begin{equation}\label{E-0+wave}
\varphi_c(0^+)= \left(f_1+f_2  \frac{\slashed P_{\!\up{D}} }{M_\up{D} }+f_3 \sl s_n+f_4 \frac{\slashed P_{\!\up{D}} \sl s_n }{M_\up{D} }   \right) \gamma^5,
\end{equation}
where the abbreviation $s_n\equiv {s_{\!\perp}}/{|\vec s\,|}$ is used; the undetermined radial wave functions $f_i~(i=1,\cdots,4)$ can be further reduced to two by the constraint condition in \eref{E-BS-HM2}, which gives the following two constraint conditions,
\begin{equation}
f_3= -\frac{|\vec s\,|(\w_1-\w_2)}{m_1\w_2+ m_2\w_1}f_1, ~~f_4= -\frac{|\vec s\,|(\w_1+\w_2)}{ m_1\w_2+m_2\w_1}f_2.
\end{equation}
Notice that the wave function of a meson with $J^{P}=0^{-}$ shares the same form with \eref{E-0+wave}.

Inserting \eref{E-0+wave} into the \eref{E-BS-HM2}, and then taking different traces, we can obtain two coupled eigenvalue equations, which can be solved numerically. The detailed procedures on solving the BSE of two-fermion systems have been presented in many previous works\,\cite{Kim2004,LiQ2016,WangT2017A,LiQ2020} and we will not copy the procedures here again. The normalization of above wave function can now be simply expressed as,
\begin{equation}
\int \frac{\d^3\vec s}{(2\pi)^3} \frac{8\w_1\w_2}{M_\up{D}(m_1\w_2+m_2\w_1)}f_1(|\vec s\,|)f_2\left(|\vec s\,|\right)=1.
\end{equation}
From this equation we know that the Salpeter wave function $f_i$ has the dimension of $\si{GeV}^{-1}$.

\subsection{\textup{Interaction kernel}}

Since the interaction kernels in a baryon and a diquark system are directly related to that in a meson, we will specify the interaction potential in a meson bound system first. As mentioned above, throughout this paper, we work under the instantaneous approximation. Then the interaction kernel of a meson based on the one-gluon exchange, reads in the momentum space, 
\begin{equation} 
iK_\up{M}(\vec t\,)=iV (\vec t\,)\gamma^\alpha \otimes \gamma_\alpha,
%V_\up{Coul}(r) = -\frac{4}{3} \frac{\alpha_s}{r}e^{-a_1r},~~~ 
\end{equation}
where $t $ denotes the momentum transfer in the interaction; and the potential $V(\vec t\,)$ in the Coulomb gauge behaves as\,\cite{Chao1992,DingYB1993,DingYB1995,Kim2004}
\begin{equation}
V(\vec t\,)= -\frac{4}{3} \frac{4\pi \alpha_s(\vec t\,)}{\vec t\,^2+a_1^2}+\left[(2\pi)^3 \delta^3(\vec t\,)\left( \frac{\lambda}{a_2}+V_0 \right)- \frac{8\pi \lambda}{(\vec t\,^2+a_2^2)^2} \right],
\end{equation}
where $\frac{4}{3}$ is the color factor; $a_{1(2)}$ is introduced to avoid the divergence in small momentum transfer zone; the kernel describing the confinement effects is introduced phenomenologically, which is characterized by the the string constant $\lambda$ and the factor $a_2$. Based on the famous Cornell potential\,\cite{Eichten1978,Eichten1980}, which behaves as the one-gluon exchange Coulomb-type potential at short distance and a linear growth confinement one at long distance, the potential used here is modified as the aforementioned one to incorporate the color screening effects\,\cite{Laermann1986,Born1989} in the linear confinement potential. $V_0$ is a free constant fixed by fitting the meson data. The strong coupling constant $\alpha_s$ has the following form, 
\[\alpha_s(\vec t\,)=\frac{12\pi}{(33-2N_f)}\frac{1}{\ln\left(a+ {\vec t\,^2}/{\Lambda^2_{\up{QCD}}}\right)},\]
where $\Lambda_\up{QCD}$ is the scale of the strong interaction, $N_f$, the active flavor, and $a=e$ is a regulator constant. 

For later convenience,  we split $V$ into two parts as
\begin{equation}
V(\vec t\,)=(2\pi)^3\delta^3 (\vec t\,) V_1+V_2 (\vec t\,),
\end{equation}
where 
\[
V_1 \equiv  \frac{\lambda}{a_2}+V_0 ,~~~V_2(\vec t\,)\equiv - \frac{8\pi \lambda}{(\vec t\,^2+a_2^2)^2} -\frac{4}{3} \frac{4\pi \alpha_s(\vec t\,)}{\vec t\,^2+a_1^2}.
\]
Namely, all the dependence on $\vec t$ is collected into $V_{2}(\vec t\,)$, while $V_{1}$ is just a constant.

The quark-antiquark pair in the meson is in the color singlet, while the quark-quark pair inside the baryon is in the color anti-triplet.  The corresponding color factors then are then $\frac{4}{3}$ and $-\frac{2}{3}$ respectively.  The interaction kernel of the diquark inside a baryon can then be expressed as, 
\begin{equation}
K_\up{D}(\vec t\,)=-\frac{1}{2} V(\vec t\,)\gamma^\alpha \otimes (\gamma_\alpha)^\up{T}.
\end{equation}
This `half rule' used here has been widely adopted in previous works involved the quark-quark bound problems\,\cite{Gershtein2000,Ebert2002,Karliner2014}. It is exact in the one-gluon exchange limit and has been validated by the measurement of the $\Xi_{cc}^{++}$ mass. Its successful extension beyond weak coupling implies that the heavy quark potential factorizes into a color dependent and a space-dependent part, with the latter being the same
for quark-quark and quark-antiquark pairs. The relative factor of $\frac{1}{2}$
then results from the color algebra, just as in the weak-coupling limit\,\cite{Karliner2017}.
Also in this work, we focus on the baryons with at least doubly heavy quarks, and then we will only consider the time component $(\alpha=0)$ of the interaction kernel, since the space components are suppressed by a factor $\frac{v}{c}$ when there is heavy quark involved in the interaction. This is also consistent with the analysis in Ref.\,\cite{Olsson1995}.

\begin{figure}[h!]
\vspace{0.5em}
\centering
\subfigure[]{\includegraphics[width=0.45\textwidth]{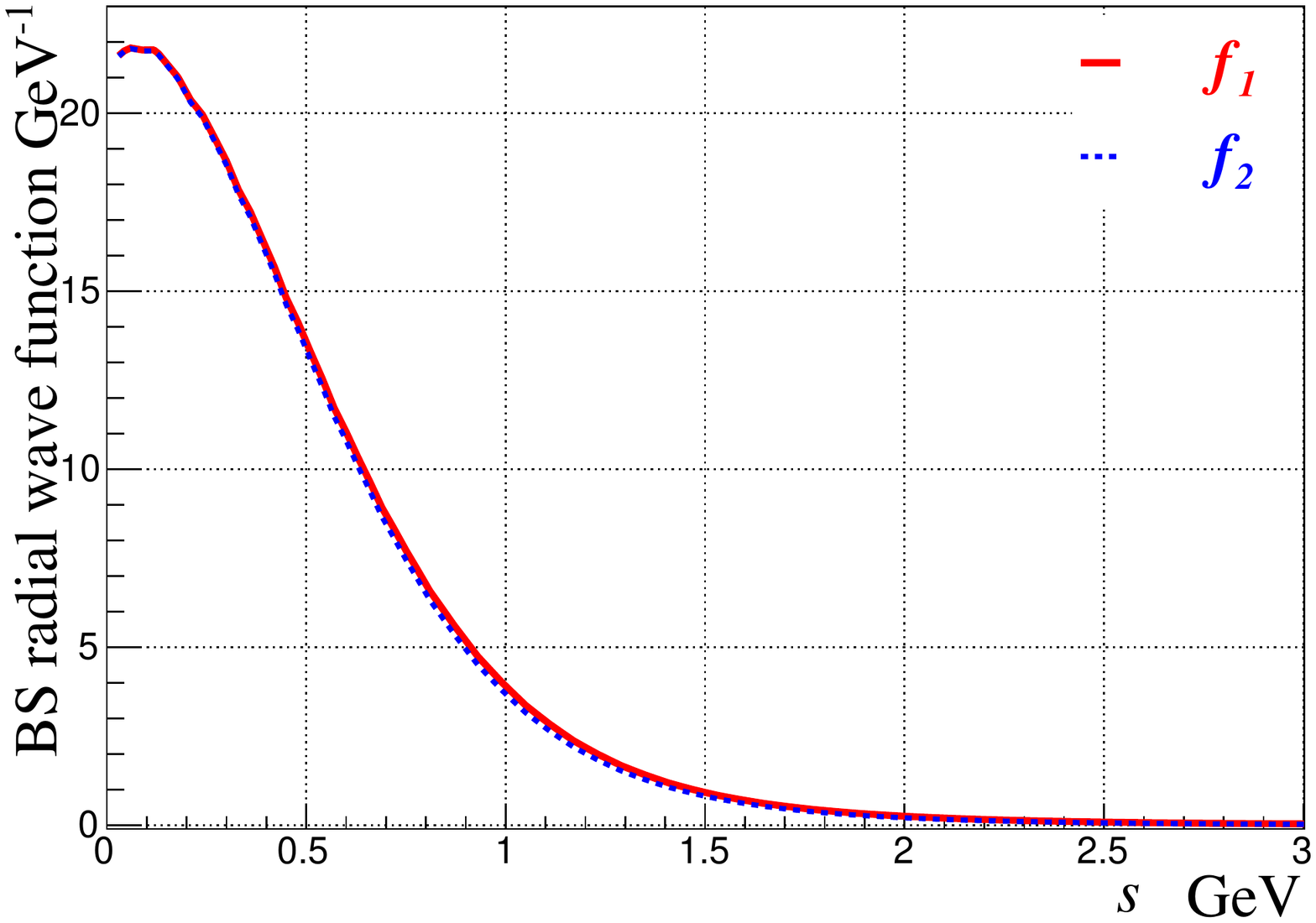} \label{Fig-bcD-n1}}
\subfigure[]{\includegraphics[width=0.45\textwidth]{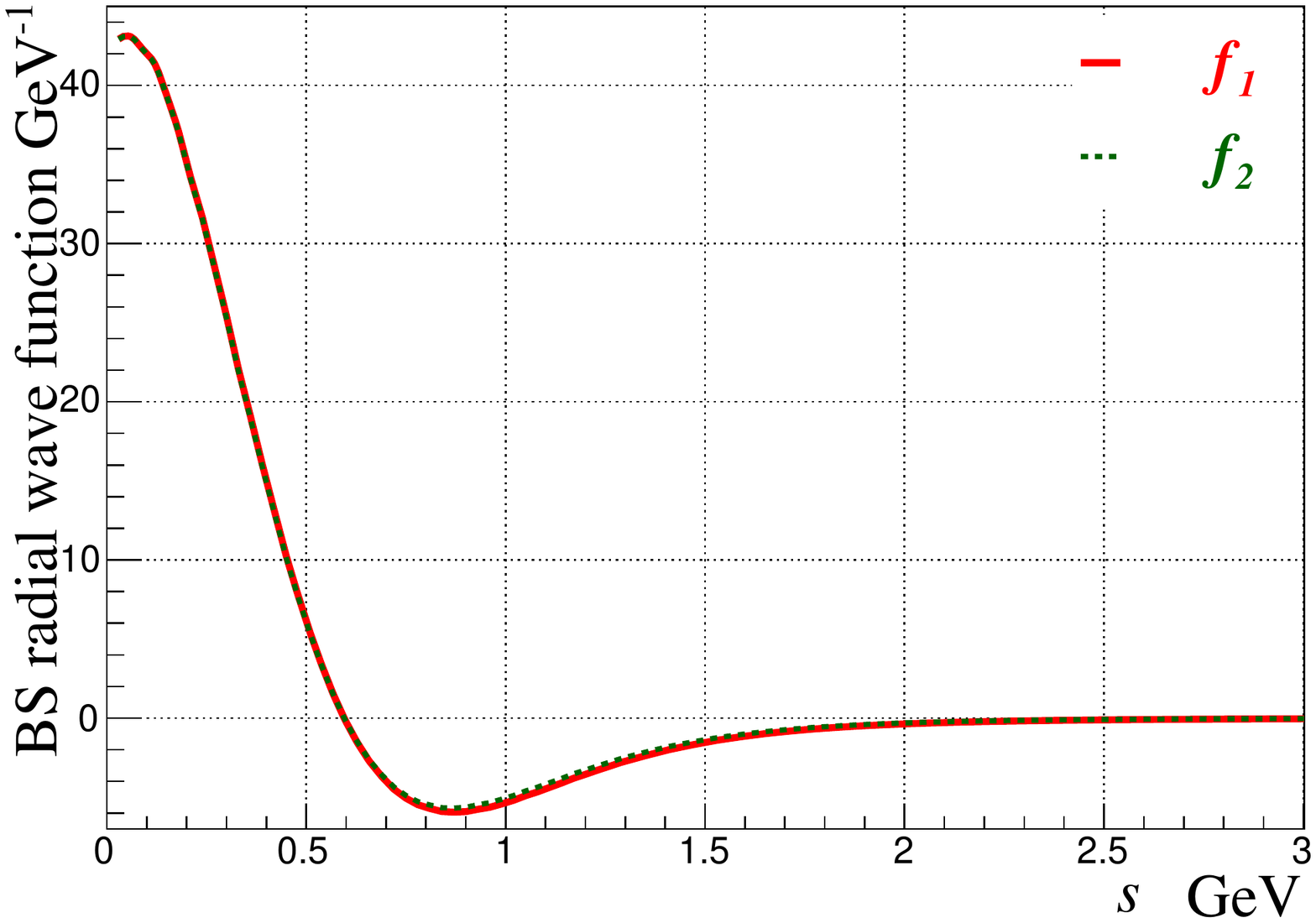} \label{Fig-bcD-n2}}\\
\caption{Radial wave functions for the $0^+$ $(bc)$-diquark in (a) the ground state and (b) the first radial excited state.}\label{Fig-wave-0+D}
\vspace{0.5em}
\end{figure}

%In this work, the parameters in the potential are fixed by fitting the mass spectra of the $1^{--}$ $c\bar c$. 
The numerical values of the model parameters used in this work are just the same with that we applied in previous calculations\,\cite{Chang2010,WangZ2012A,WangT2013,WangT2013A,LiQ2016,LiQ2017,LiQ2017A,LiQ2020},  and determined by fitting to the corresponding mesons
 namely,
\[
a =e=2.7183,~~~ \lambda =0.21~\si{GeV}^2, ~~~ \Lambda_\text{QCD} =0.27~\si{GeV},  ~~~a_1 =a_2=0.06~\si{GeV},
\]
and the constitute quark masses used are
\[
m_u =0.305~\si{GeV},~~ m_d =0.311~\si{GeV}, ~~ m_s =0.5~\si{GeV}, ~~ m_c =1.62~\si{GeV}, ~~m_b=4.96~\si{GeV}.
\]
The free parameter $V_0$ is fixed by fitting the meson mass $m_{B_c}$ to the experimental value. In this work, we found $V_0=-0.21~\si{GeV}$ for the $0^+$ $(bc)$-diquark.  Finally, the corresponding $(bc)$-diquark mass are $M_{(bc)}(1S)=6.563\,\si{GeV}$ and $M_{(bc)}(2S)=6.892\,\si{GeV}$ for the ground and the first radially excited states respectively, and the corresponding radial wave functions are showed in \autoref{Fig-wave-0+D}.

%\begin{table}[h!]
%\caption{Mass spectra of the $0^+$ $(bc)$ systems in units of GeV.}\label{Tab-Mass-bcD}
%\vspace{0.2em}\centering
%\begin{tabular}{ cccccccccc }
%\toprule[2pt]
%$n_dL_d$	&$1S$				&$2S$				&$3S$				&$4S$\\%			&$5S$			&$6S$			&$7S$			&$8S$\\
%\midrule[1.5pt]
%$(bc)$ 		&$6.563$			&$6.892$			&$7.111$			&$7.279$	\\%	&$7.417$		&$7.533$		&$7.633$	 	&$7.720$\\
%\bottomrule[2pt]
%\end{tabular}
%\end{table}

The $0^+$ diquark wave function at zero point is helpful in many applications, which is defined as
\begin{gather}
\left. \varphi(\vec x)\right|_{\vec x=0} = \int \frac{\up d^3 \vec s}{(2\pi)^3} \varphi_c(\vec s) =\left( \psi_\up{P} + \psi_\up{A} \frac{\slashed P_\up{D}}{M_\up{D}} \right) \gamma^5,
\end{gather}
where $\psi_\up{P}$ and $\psi_\up{A}$ are calculated from the Salpeter radial wave functions,
 \begin{gather}
\psi_\up{P} =  \int \frac{\up d^3 \vec s}{(2\pi)^3} f_1,~~
\psi_\up{A} = \int \frac{\up d^3 \vec s}{(2\pi)^3} f_2.
\end{gather}
The obtained numerical results are $\psi_\up{P}=0.311 $  and $\psi_\up{A}=0.274\,\si{GeV}^2$ for the ground state; $\psi_\up{P}=0.273$ and $\psi_\up{A}=0.230\,\si{GeV}^2$ for the first excited state.

\subsection{\textup{The diquark form factors and wave function at zero point}}
\begin{figure}[h!]
\centering
\subfigure[Gluon couples with quark-2 in the diquark.]   {\includegraphics[width=0.48\textwidth]{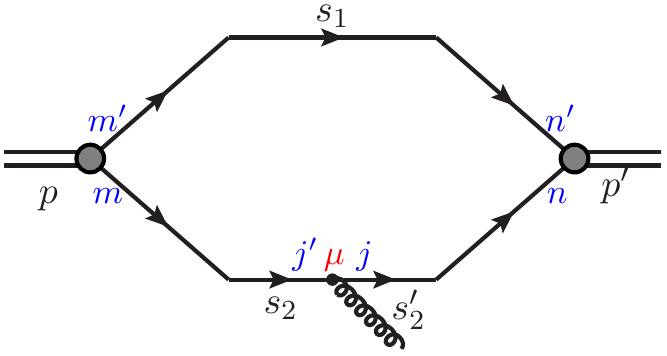} \label{Fig-Form-D0-A}}
\subfigure[Gluon couples with quark-1 in the diquark.]   {\includegraphics[width=0.48\textwidth]{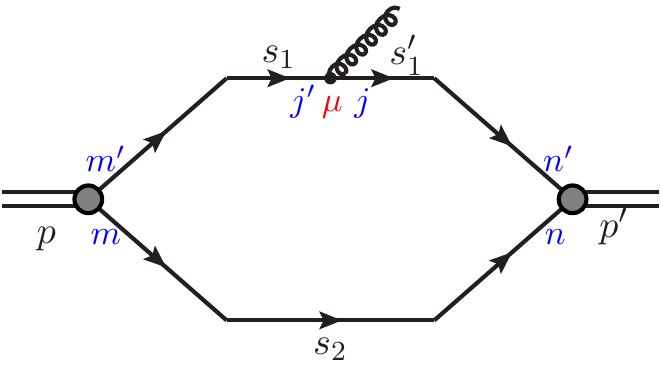} \label{Fig-Form-D0-B}}
\caption{Scattering vertex between the diquark and a gluon.} \label{Fig-Form-D0}
\end{figure}
The Feynman diagrams of the doubly heavy diquark coupling to a gluon are showed in \autoref{Fig-Form-D0}. The variables involved are defined as: $M_\up{1}$ denotes the diquark effective mass; $p~(p\rq{})$, the diquark momenta before (after) coupling to the gluon. The internal momenta $s'$ in the final state is defined as $s^{\prime}=\lambda^{\prime}_2s^{\prime}_1-\lambda^{\prime}_1s^{\prime}_2$, 
where $\lambda'_i \equiv \frac{m'_i}{m_1'+m_2'}=\lambda_i$ since $m_i^{\prime}=m_i$. Also in \autoref{Fig-Form-D0-A}$ ~s'_1=s_1$, then $s$ and $s'$ are related by $s'=\lambda_1(p-p')+s$.

Considering the the Lorentz conditions and the vector current conservation, the form factor $\mathcal{A}^{\mu}$ of the $0^+\to 0^+$ diquark coupling a a gluon can generally be expressed as,
\begin{equation}
\mathcal{A}^{\mu}=C_{00} (p+p')^\mu,% \\
%&\overset{\up{IA}}{\simeq}  \sigma_1(t) g^{\alpha\beta} (p_1 + k_1)^\mu,
\end{equation}
where $C_{00}$ here are explicitly dependent on $\Delta p^2$ with $\Delta p\equiv p-p'$ denoting the momentum transfer. 
On the other hand, according to the Feynman diagram \autoref{Fig-Form-D0}, the form factor $\mathcal{A}^\mu$ can be directly expressed by the transition amplitude,
\begin{equation}
\begin{aligned}
\mathcal{A}^{\mu}=\frac{1}{2}\left(\mathcal{A}_1+\mathcal{A}_2 \right),
\end{aligned}
\end{equation}
where the factor $\frac{1}{2}$ comes from the normalization convention; the amplitude  $\mathcal{A}_{1(2)}$ corresponds to the process when the gluon interacts with the $b(c)$-quark. $\mathcal{A}_1$ is explicitly described by the BS vertex as,
\begin{equation} \label{E-formD0D0-4D}
\begin{aligned}
\mathcal{A}_1&=-\int \frac{\d^4 s}{(2\pi)^4} \up{tr}~\bar{\Gamma}_c(p',s') S(s_1)\Gamma_c(p,s) S(-s_2) \gamma^\mu S(-s'_2)
%&\simeq \int \frac{\up{D}^3 \vec s}{(2\pi)^3} \up{Tr}~\bar{\varphi}_c(p',s'_{\!\perp})\gamma^0 \varphi_c(k_1,s_{\!\perp}) \gamma^\mu,
&=C^{(1)}_{00}(p+p')^\mu.
\end{aligned}
\end{equation}
By performing the contour integration over $s^0$, $\mathcal{A}_1$ can be further expressed by the three-dimensional integral of the Salpeter wave functions $\varphi_c$. Also it is easy to obtain $\mathcal{A}_2$ by interchange $m_1$ and $m_2$, namely,  
\[ \mathcal{A}_2=C^{(2)}_{00}(p+p')^\mu =\mathcal{A}_1(m_1\rightleftharpoons m_2),  \]
and then we obtain the form factor as $C_{00}=\frac{1}{2}\left[C^{(1)}_{00}+C^{(2)}_{00}\right]$.
\begin{figure}[h!]
\centering
\subfigure[]   {\includegraphics[width=0.48\textwidth]{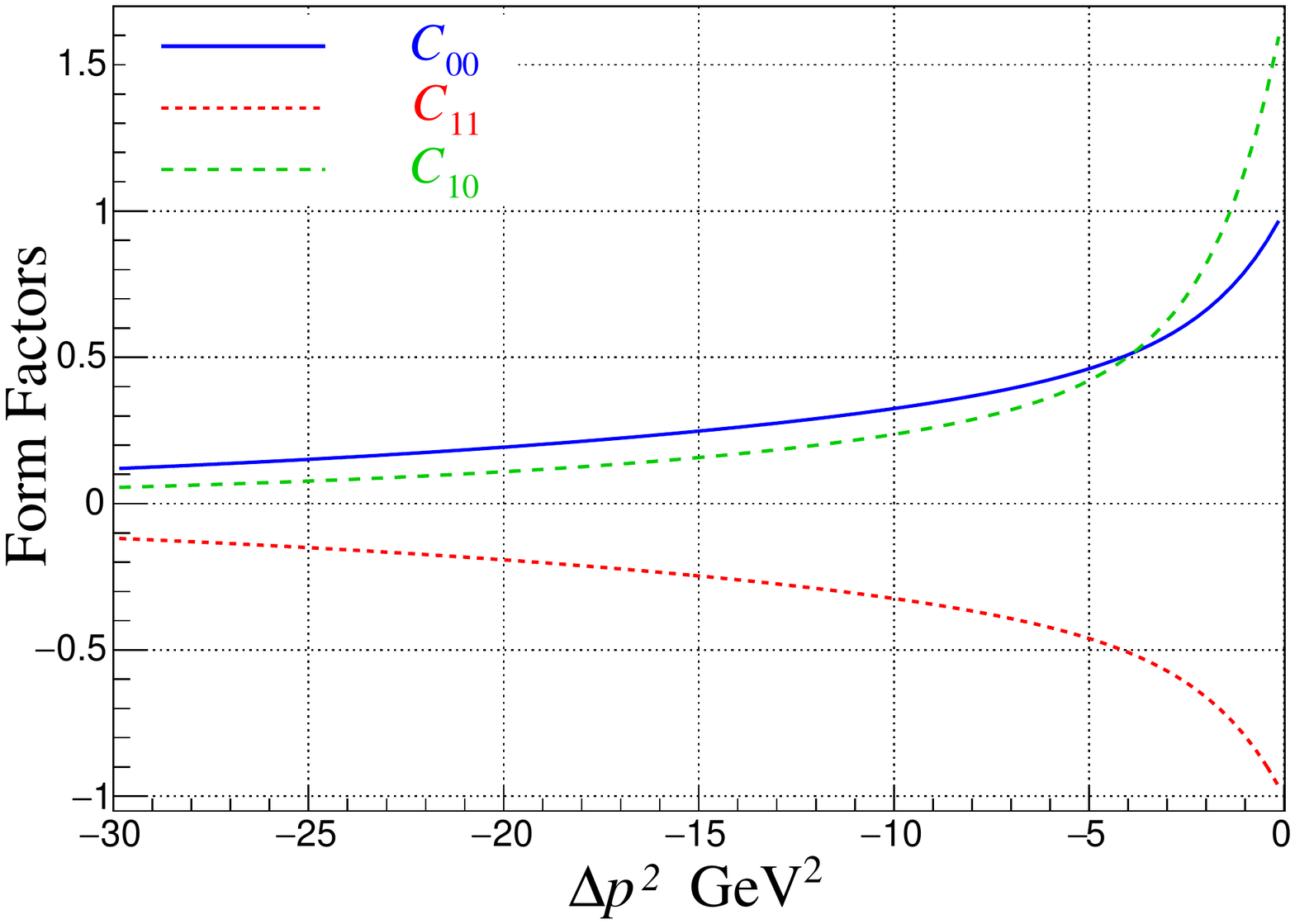} \label{Fig-Form-D0-A}}
\subfigure[]   {\includegraphics[width=0.48\textwidth]{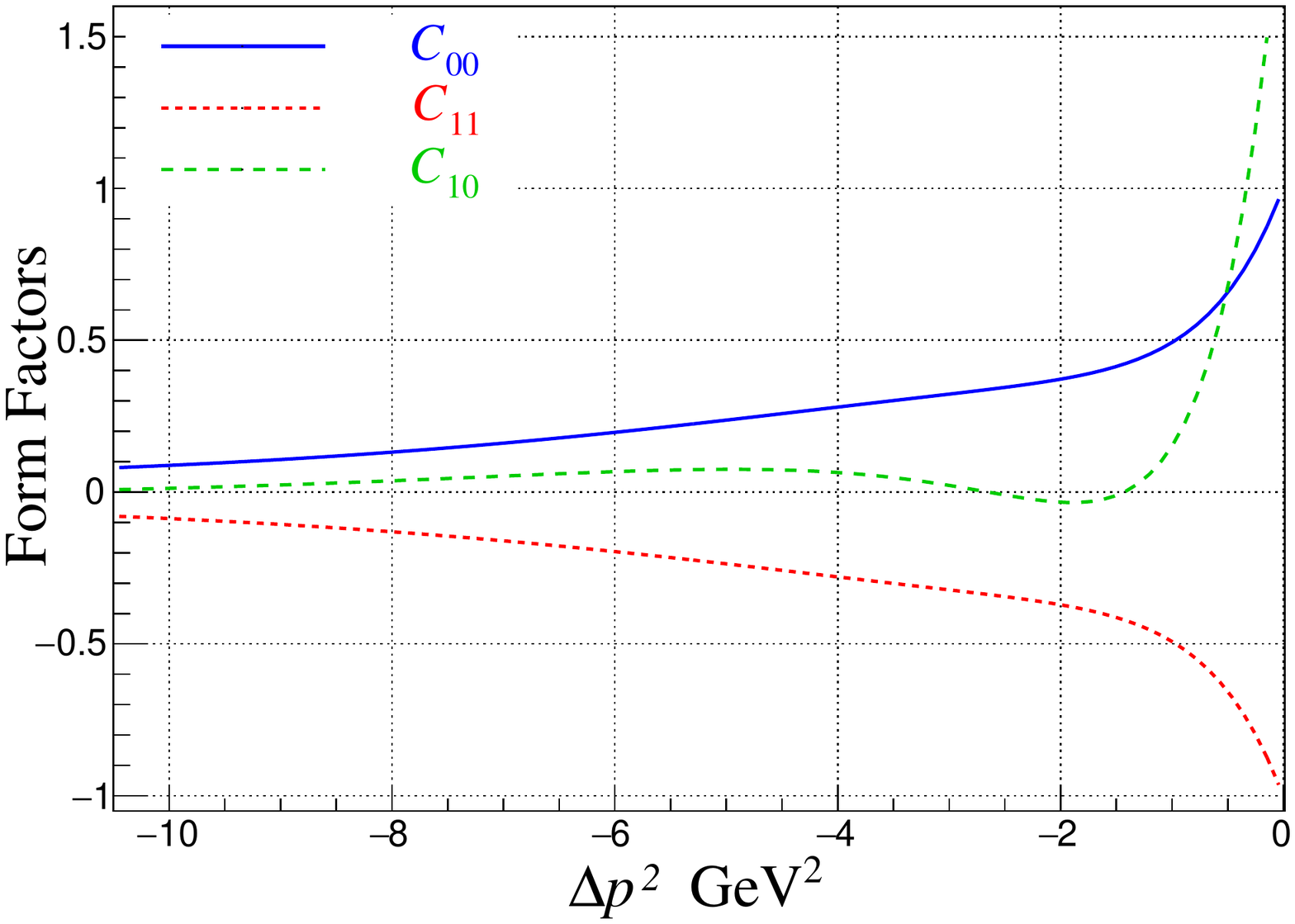} \label{Fig-Form-D0-B}}
\caption{The obtained form factors $C_{11}$, $C_{00}$, and $C_{10}$ of the diquark in  (a) the ground state, and (b) the first excited state.} \label{Fig-Xibc-formfactors}
\end{figure}

Inserting the $0^+$ Salpeter wave function, namely, \eref{E-0+wave}, into above equation, by calculating the trace and the integration, we get the numerical values of the diquark form factors, which are graphically shown in \autoref{Fig-Xibc-formfactors}.
%The form factor of $cg\to cg$ in the ground diquark can be parameterized as,
%\begin{equation}
%\sigma(t^2)=A_1e^{\kappa_1t^2}+A_2e^{\kappa_2t^2}+A_3e^{\kappa_3t^2},
%\end{equation}
%where the numerical parameters  we obtained are $A_1=0.4, ~\kappa_1=0.293, ~A_2=0.573,~\kappa_2=0.614$, $A_3=0.027,~\kappa_3=0.087$.
%\begin{gather}
%\sigma(t^2)=A_1e^{\kappa_1t^2}+A_2e^{\kappa_2t^2}+(1-A_1-A_2)e^{\kappa_3t^2},
%\end{gather}
%where the numerical parameters  we achieved are $A_1=3.90\times10^{-2}, ~\kappa_1=5.92\times10^{-2}, ~A_2=4.43\times10^{-1}$, $\kappa_2=3.84\times10^{-1}$, $\kappa_3=2.14\times10^{-1}$.
%\begin{figure}[htpb]
%\centering
%\includegraphics[width = 0.60\textwidth, angle=0]{formAll}
%\caption{The obtained form factors $C_{11}$, $C_{00}$, and $C_{10}$.}
%\label{Fig-Xibc-formfactors}
%\end{figure}

The form factor describing $1^+$ to $0^+$ (or vice versa) diquark transition is directly related to the mixing effects between the $\frac{1}{2}^+$ baryon doublet $\Xi_{bc}$ and $\Xi'_{bc}$, or $\Omega_{bc}$ and $\Omega'_{bc}$. Note this transition could only happen through the $P$-wave gluon exchange. %As mentioned before, when the $(bc)$-diquark is in the $1^+\,(0^+)$ spin configuration, the flavor wave function would be symmetric\,(anti-symmetric), namely, $\psi(\up{flavor})=\frac{1}{\sqrt{2}}(\ket{bc} \pm \ket{cb})$. 
By a similar analysis, the form factor for ${1^+}$ to ${0^+}$  diquark transition can be expressed as,
\begin{gather}
\mathcal{A}^{\alpha\mu}=\frac{1}{2}\left(\mathcal{A}^{\alpha\mu}_1+ \mathcal{A}^{\alpha\mu}_2 \right) =i \frac{2E_{\!F}}{M_{(0)} M_{(1)} } C_{10} \epsilon^{\alpha\mu pp'},
\end{gather}
where $\epsilon^{\alpha\mu pp'}=\epsilon^{\alpha\mu \nu \beta}p_\nu p'_\beta$ and $\epsilon^{\alpha\mu \nu \beta}$ is the Levi-Civita antisymmetric tensor; the extra Lorentz index $\alpha$ comes from the vector wave function of the initial $1^+$ diquark; $M_{(0)}$ and $M_{(1)}$ are the $0^+$ and $1^+$ diquark masses respectively; the factor $ E_F $ here helps keep the dimensions of the $C_{ij}$ consistent and $E_{F}=({M^2_F+\vec p\,'^2})^{\frac12}$ with $M_F$ denoting the constituent mass of the diquark after transition, which is $M_{(0)}$ in the $1^+\to 0^+$ transition and is $M_{(1)}$ in the $0^+\to 1^+$ transition. Notice flavor-symmetric $\psi(\up{flavor})$ of the diquark becomes antisymmetric after the transition. The Feynman amplitude $\mathcal{A}^{\alpha\mu}_1$ is explicitly expressed by the BS vertex as,
%\begin{equation}
%\begin{aligned}
%\Sigma^{\alpha\mu}=\frac{1}{2}\left(\Sigma^{\alpha\mu}_1-\Sigma^{\alpha\mu}_2 \right),
%\end{aligned}
%\end{equation}
%where the amplitude $\Sigma^{\alpha\mu}_1$ is explicitly expressed as,
\begin{equation}
\begin{aligned} \label{E-formD1D0-4D}
\mathcal{A}^{\alpha\mu}_1&=-\int \frac{\d^4 s}{(2\pi)^4} \up{tr}~\bar{\Gamma}_c(p',s') S(s_1)\Gamma^{\alpha}_c(p,s) S(-s_2) \gamma^\mu S(-s'_2)=i\frac{2E_F}{M_{(0)} M_{(1)}} C^{(1)}_{10} \epsilon^{\alpha\mu pp'},
\end{aligned}
\end{equation}
where the $1^+$ Salpeter wave function had been given in our previous work\,\cite{LiQ2020}.
And again we have $\mathcal{A}^{\alpha\mu}_2 = \mathcal{A}^{\alpha\mu}_1(m_1\rightleftharpoons m_2)$. Then the form factor $C_{10}=\frac{1}{2}[C^{(1)}_{10}+C_{10}^{(2)}]$ can be obtained numerically by a similar procedure as above, and the obtained numerical result is also displayed in \autoref{Fig-Xibc-formfactors}.  The form factor $\mathcal{A}^{\alpha\mu}$ for the $0^+$ to $1^+$ $(bc)$-diquark  transition is just the same with that in the $1^+\to 0^+$ case.

%Besides the $1^+$ to $0^+$ $(bc)$-diquark transition inside the $\frac{1}{2}^+$ $(bcq)$ baryon states, there could exist the opposite transition. Since this form factor, namely the $0^+\to 1^+$ $(bc)$-diquark  transition is just the same with that in the $1^+\to 0^+$ case, we will not repeat the calculations again and just give a brief proof in the appendix\ref{A-1}.

\section{$\frac{1}{2}^+$ baryons consisting of the $0^+$ doubly heavy diquark core and a light quark}\label{Sec-3}

\begin{figure}[h!]
\centering
\includegraphics[width = 0.9\textwidth, angle=0]{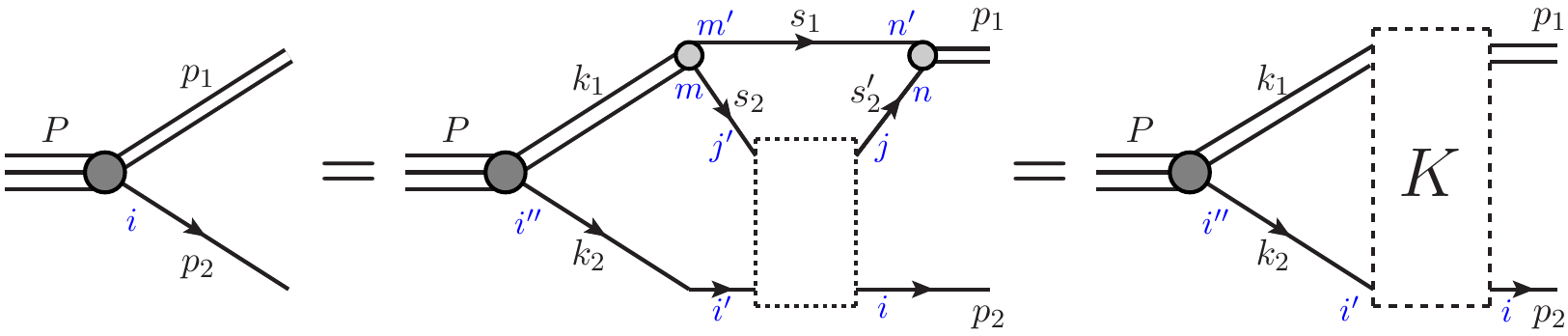}
\caption{Bethe-Salpeter equation of the baryon based on the diquark-quark picture. $P$ denotes the baryon momentum, $p_1(k_1)$, the diquark momentum, and $p_2(k_2)$, the third quark momentum ; The Romans (blue) denote the Dirac indices.}\label{Fig-BS-B-D0}
\end{figure}

In this section, we first give the Bethe-Salpeter equation of the doubly heavy baryon as the bound state of the $0^+$ heavy diquark core and the remaining light quark; then in the instantaneous approximation, we derive the corresponding three-dimensional Salpeter equation; finally, the baryon Salpeter wave functions are constructed and solved to obtain the $J^P=\frac{1}{2}^+$ $(bcq)$ baryons.

\subsection{\textup{BSE of baryons with $0^+$ heavy diquark cores}}
The baryon BSE based on the diquark-quark picture is showed schematically in \autoref{Fig-BS-B-D0}, which is expressed by the matrix notation as,
\begin{equation}\label{E-BS-BD0}
\Gamma(P,q,\xi)=\int  \frac{\d^4 k}{(2\pi)^4} (-i)K(p_1,k_1;p_2,k_2) \left[S(k_2)\Gamma(P,k,\xi) D(k_1)\right],
\end{equation}
where $(-i)K(p_1,k_1;p_2,k_2)$ represents the effective quark-diquark interaction kernel, which depends on the diquark and the third quark's momenta $p_1,~k_1$ and $p_2,~k_2$ respectively; $\Gamma(P,q,\xi)$, the baryon vertex; $P$, the baryon momentum, $P^2=M^2$ with $M$ denoting the baryon mass; $r$, the baryon polarization state; $S(k_2)$, the Dirac propagator of quark with momentum $k_2$; $D(k_1)$ is the scalar propagator; the baryon wave function can be defined as usual 
\begin{equation} \label{E-Salpeter-wave}
B(P,q,\xi)=S(p_2) \Gamma(P,q,\xi) D(p_1) .
\end{equation}
%Then the BS vertex in \eref{E-BS-BD0} can be simplified as the integration over $B(P,q)$.
The symbols $P$ and $\xi$ in the BS vertex $\Gamma(P,q,\xi)$ and wave function $B(P,q,\xi)$ would be temporarily omitted unless it is necessary to write them explicitly.
The two internal momenta $q$ and $k$ in the baryon BSE are defined as usual,
\[
q=\alpha_2 p_1 - \alpha_1 p_2,~~~ k=\alpha_2 k_1 - \alpha_1 k_2,
\]
where $\alpha_i\equiv \frac{m_i}{m_1+m_2}$; $m_1$ denotes the diquark effective mass, and $m_2$, the third quark constituent mass.
With the momentum conservation, by simple analysis, there are only 4 independent variables here, which can be chosen as, the baryon total momentum $P$, quark-diquark internal momentum $k$ and $q$, and the internal momentum of the diquark $s$. Then other variables can be expressed by these four variables. We point out that the Bethe-Salpeter equation given here can also be applied to deal with the pentaquark states formed by a (pseudo)scalar meson and a baryon\,\cite{XuH2020}.

The effective interaction kernel $(-i)K(p_1,k_1;p_2,k_2)$ can be considered as the diquark-quark scattering, which is expressed by the BS vertex as,
\begin{equation}
\begin{aligned}\label{E-Kb}
  &(-i)K_{ii'}(p_1,k_1;p_2,k_2) \\
=& iV (k-q) \gamma^\mu_{ii'} \frac{1}{2}\int \frac{\d^4 s}{(2\pi)^4} \up{tr}~ \bar\Gamma_\up{c}(p_1,s') S(s_1) \Gamma_\up{c}(k_1,s) S(-s_2) \gamma_\mu S(-s'_2)+(m_1\rightleftharpoons m_2).
\end{aligned}
\end{equation}
Notice that the integration part in \eref{E-Kb} above is just the $0^+$ diquark form factor which we have obtained in previous section, only the on-shell $p,~p'$ replaced by the off-shell momenta $k_1$ and $p_1$ respectively.  Here the on-shell form factors are generalized as an effective description on the coupling of the diquark to a gluon; besides, we do not consider the self-energy correction of the doubly heavy $(bc)$-diquark. Then the interaction kernel $(-iK)$ can now be written as a simple expression,
\begin{equation}\label{E-Kb3}
(-i)K(p_1,k_1;p_2,k_2)=- iV (k-q)\mathcal{A}_{\mu} \gamma^\mu ,
\end{equation}
where $\mathcal{A}_{\mu}=C_{00} (p_1+k_1)_\mu$ is the obtained $0^+$ $(bc)$-diquark form factor. Notice that now the interaction potential between the diquark and the light quark is smeared by the diquark form factor $C_{00}$. As mentioned above, we only concentrate on the dominate time component of this vector interaction. 

Now we move to obtain the three-dimensional Salpeter equation\,(SE) of the baryons. First,  we define the baryon Salpeter wave function as usual  $\varphi(q_\perp) \equiv -i\int \frac{\up d q_P}{2\pi} B(q)$, where $q_P=\hat P\cdot q$, and $q_\perp=q-q_P \hat P$ with $\hat P\equiv \frac{P}{M}$. On the other hand, under the instantaneous approximation, the interaction potential $V (k-q)\sim  V (k_\perp - q_\perp)$, which is equivalent to $q^0=k^0$, and thus we have $\varrho\equiv (k_1+p_1)^0=2(\alpha_1M+q_P)$. Notice the factor $\varrho$ plays an important role in the derivation of the three-dimensional SE. Now the BS vertex in \eref{E-BS-BD0} can be expressed as $\Gamma(q)=\varrho \gamma^0 \Theta(q_\perp)$, where $\Theta$ is a three-dimensional integral over $k_\perp$, namely,
\begin{equation} \label{E-vertex1}
\Theta (q_\perp) \equiv \int \frac{\up d^3 k_\perp}{(2\pi)^3} \varkappa (k_\perp-q_\perp) \varphi(k_\perp),
\end{equation}
where the instantaneous kernel $\varkappa(x) \equiv C_{00} V(x)$. Notice now both $\varkappa$ and $\Theta$ does not explicitly depend on the $q_P$ nor $M$.

Performing a contour integral over $q_P$ on both sides of  \eref{E-Salpeter-wave}, we obtain the following three-dimensional (Bethe-)Salpeter equation (see appendix \ref{A-Salpeter}),
\begin{equation} \label{E-BSB-D0-2}
M\varphi %=\mathcal{H}\varphi
=(w_1+w_2) H_2( p_{2\perp})\varphi + {H_2}(p_{2\perp})  \Theta  ,
\end{equation}
where $w_i=(M_i^2+\vec p_i^{\,2})^{\frac12}$ with $M_1$ and $M_2$ denoting the masses of diquark and the third quark respectively; $H_2(p_{2\perp})=\frac{1}{w_2}(p_{2\perp}^{\alpha} \gamma_\alpha +M_2)\gamma_0$; the bound state mass $M$ behaves as the eigenvalue of above baryon Salpeter equation.
The meaning of this equation is quite obvious. The baryon mass has two contributions, the kinetic energy part and the potential energy part.
The normalization of the baryon wave function behaves as,
\begin{equation}\label{E-Norm-D0}
\int \frac{\d^3 q_\perp}{(2\pi)^3} \frac1M\bar\varphi(P,q_\perp,\xi')   O \varphi(P,q_\perp,\xi)=\delta_{\xi \xi'},
\end{equation}
where the operator $O$ is defined as $ O \equiv \gamma^0\left( \alpha_1 M  H_2 +  w_q \right) $ with $ w_q \equiv\alpha_2w_1-\alpha_1w_2$, and the more detailed derivations can be found in appendix \ref{A-Norm}.
%
%The normalization form inspires us to define a scalar product of the Salpeter wave functions $\varphi_j(P,q, \bar r)$ and $\varphi_i(P,q,r)$ as follow
%\begin{gather}
%\langle \varphi_j(q_\perp, \bar r)| \varphi_i(q_\perp, r)\rangle = \int \frac{\up{D}^3 q_\perp}{(2\pi)^3} \bar\psi(P,q_\perp,\bar r) \gamma^0 \left(\alpha_1 \hat H( p_{2\perp})+\frac{\w_q}{M_i} \right) \varphi(P,q_\perp,r),
%\end{gather}
%where both $\varphi_j(P,q, \bar r)$ and $\varphi_i(P,q,r)$ are the normalized Salpeter wave functions, which fulfill the \eref{E-BSB-D0-2}, namely, 
%\begin{gather*}
%M_i \varphi_i(P,q_\perp, r)=\mathcal{H} \varphi_i(P,q_\perp,r),~~~
%M_j \varphi_j(P,q_\perp, \bar r)=\mathcal{H} \varphi_j(P,q_\perp,\bar r).
%\end{gather*}
%Then it is easy to find that the normalization condition of the Salpeter wave function can be expressed as $\langle\varphi (P,q_\perp, \bar r)| \varphi(P,q_\perp, r)\rangle=\delta_{r\bar r}$.

\subsection{\textup{Baryon Salpeter wave functions with $J^P=\frac{1}{2}^+$}}

In the ground state, namely, the angular momentum $L=0$, the $0^+$ $(bc)$-diquark and a light quark $u,d$, or $ s$ can form the $J^P=\frac{1}{2}^+$ baryon states $\Xi'^+_{bc}$, $\Xi'^0_{bc}$, or $\Omega'_{bc}$, which can be described by the following Salpeter wave function,
\begin{gather}\label{E-wave-D0-1H}
\varphi(P,q_\perp,\xi)=A(q_\perp) u(P,\xi),~~~A(q_\perp)\equiv h_1 + h_2 \slashed y,
\end{gather} 
where the new abbreviation $y\equiv \frac{q_\perp}{|\vec q\,|}$ is used; the radial wave functions $h_i(|\vec q\,|)~(i=1,2)$ are just explicitly dependent on $|\vec q\,|$; $u(P,\xi)$ is the Dirac spinor with polarization state indicated by $\xi$. %In terms of the spherical harmonics $Y_l^m$, the wave function can be rewritten as,
%\begin{gather} \label{E-wave0-Ylm}
%\varphi(P,q_\perp,r)= C_0 \left[h_1 Y_0^0  + C_1 h_2 \left( Y_1^{+1}\gamma^- + Y_1^{-1} \gamma^+ -Y_1^0 \gamma^3\right)   \right] u(P,r),
%\end{gather}
%where $C_0=2 \sqrt{\pi}$ and $C_1={1}/{\sqrt{3}}$; $\gamma^{\pm} = \mp (\gamma^1\pm i \gamma^2) {\sqrt{2}}$. 
It is clear to see that $h_1$ and $h_2$ parts represent the $S$ and $P$-wave components, respectively.

Inserting the $\frac{1}{2}^+$ Salpeter wave function into the \eref{E-Norm-D0}, summing over the polarization states $\xi$, then we obtain the following specific normalization condition, 
\begin{equation}
%&\frac{1}{(2\times \frac{1}{2}+1)\times 2M}\int \frac{\up{D}^3 q_\perp}{(2\pi)^3} \bar\varphi\gamma^0 2\left(\alpha_1 M\hat H( p_{2\perp})+\w_q \right) \varphi(P,q_\perp) \\
%=&\frac{1}{2M}\int \frac{\up{D}^3 q_\perp}{(2\pi)^3} \bar u(P,r)\bar A\gamma^0 \left( \alpha_1 M\hat H( p_{2\perp})+ \w_q \right) A(q_\perp) u(P,r) \\
\int \frac{\d^3 q_\perp}{(2\pi)^3} 2\left[ w_q(h_1^2+h_2^2)+\alpha_1M\frac{1}{w_2}(M_2h_1^2-2qh_1h_2-M_2h_2^2) \right]=1.
\end{equation}
Then inserting this wave function into the obtained Salpeter \eref{E-BSB-D0-2}, 
\begin{equation} \label{E-eigen-D1-1H}
MAu(P,\xi)=(w_1+w_2) H_2( p_{2\perp})A u(P,\xi) + {H}_2(p_{2\perp})\int \frac{\up d^3 k_\perp}{(2\pi)^3} \varkappa(k_\perp-q_\perp) A(k_\perp)u(P,\xi).
\end{equation}
Multiplied by $\bar u(P,\xi)$ and then summing over the polarization states, we can eliminate the spinor in above equation.  Next by taking different traces, we obtain 2 coupled eigenvalue equations on the baryon mass $M$,
\begin{equation}
\begin{aligned}
Mh_1(\vec q\,)&=+E_1 h_1(\vec q\,)-E_2 h_2(\vec q\,)+\frac{1}{w_2} \int \frac{\up d^3 \vec k}{(2\pi)^3} \mathcal{V}_2(\vec q-\vec k) \left[m_2 h_1(\vec k) -q\cos\theta h_2(\vec k)\right],\\
Mh_2(\vec q\,)&=-E_2 h_1(\vec q\,)-E_1h_2(\vec q\,) -\frac{1}{w_2} \int \frac{\up d^3 \vec k}{(2\pi)^3} \mathcal{V}_2 (\vec q-\vec k) \left[q h_1(\vec k) +m_2\cos\theta h_2(\vec k)\right],
\end{aligned}
\end{equation}
where $E_1=\frac{M_2}{w_2}(w_1+w_2+\mathcal{V}_1)$, $E_2=\frac{q}{M_2}E_1$; $\mathcal{V}_i\equiv C_{00} V_{i}$; $\theta$ denotes the angle between $\vec k$ and $\vec q$.
Solving these eigen equations numerically, we can obtain the mass spectra and the corresponding wave functions.

\section{Bethe-Salpeter equation of the $(bcq)$ baryons incorporating the mixing effects}\label{Sec-4}

Generally, the $(bc)$-diquark core in $\Xi_{bc}$ or $\Omega_{bc}$ is in the mixing state of $0^+$ and $1^+$. To obtain the real physical baryon states, the inner transition between the axialvector and the scalar $(bc)$-diquark states should be incorporated in solving the BSE. This inner transitions then connect the flavor symmetric $(bcq)$ baryon states with the flavor antisymmetric ones.

\subsection{Bethe-Salpeter equation considering the mixing effects}
\begin{figure}[h!]
\centering
\subfigure[]{\includegraphics[width=0.8\textwidth]{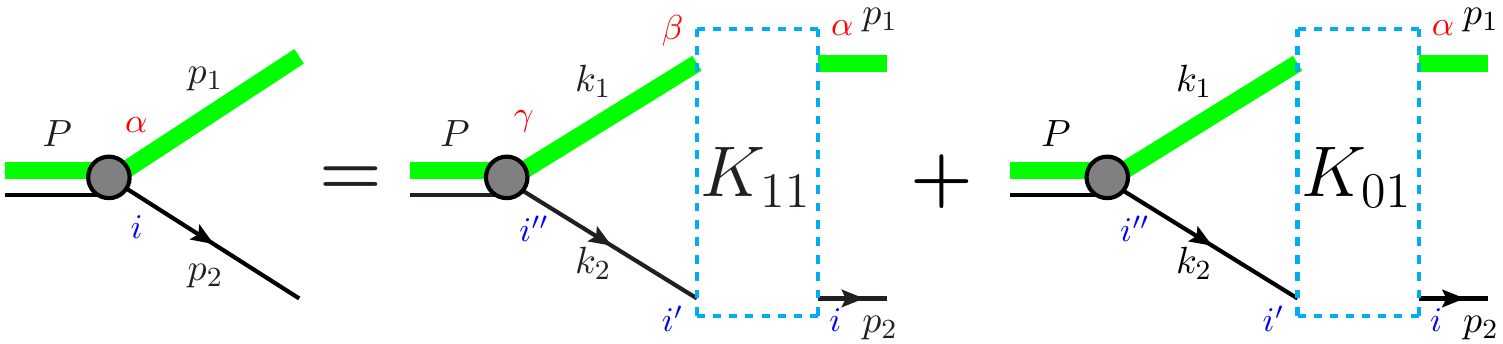} \label{Fig-BSE-0-1}}
\subfigure[]{\includegraphics[width=0.8\textwidth]{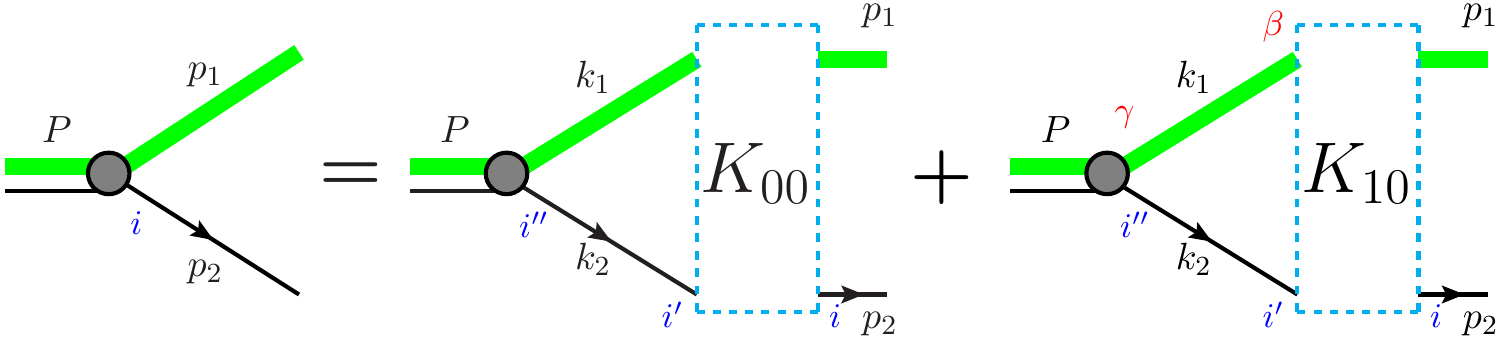} \label{Fig-BSE-1-0}}
\caption{Bethe-Salpeter equations of the baryons with different flavors consist of two coupled equations in the diquark picture.  The Greeks~(red) are used for the Lorentz indices; the Romans (blue), the Dirac indices. $P,~p_1(k_1),~p_2(k_2)$ denote the momenta of the baryon, constituent diquark, and the light quark respectively; $K_{mn}\,(m=1,0)$ denotes the effective interaction kernel between the diquark and the third quark in a baryon.}\label{Fig-BS-B}
\end{figure}

The Bethe-Salpeter equations of the doubly heavy baryons with different flavors are schematically depicted in \autoref{Fig-BS-B}. Notice in \autoref{Fig-BSE-0-1} there is an extra Feynman diagram compared with that of the baryons with pure $1^+$ $(bc)$-diquark, namely the ``$K_{01}$'' part; and in \autoref{Fig-BSE-1-0} there is the extra ``$K_{10}$'' Feynman diagram compared with that of the baryons consisting of the pure $0^+$ $(bc)$-diquark core. The corresponding BSEs are expressed as the four-dimensional integral of the inner relative momentum $k$, 
\begin{align}
\Gamma^\alpha(q) &=\int \frac{\up d^4 k}{(2\pi)^4} (-i)K_{11}^{\alpha\beta}[S(k_2) \Gamma^\gamma(k) D_{\beta\gamma}(k_1)]+ (-i)K_{01}^{\alpha}[S(k_2) \Gamma(k) D(k_1)], \label{E-BSE-A1} \\
\Gamma(q) &=\int \frac{\up d^4 k}{(2\pi)^4} (-i)K^{\beta}_{10}[S(k_2) \Gamma^\gamma(k) D_{\beta\gamma}(k_1)] + (-i)K_{00} [S(k_2) \Gamma(k) D(k_1)]. \label{E-BSE-A0}
\end{align}
Before moving on, we specify the symbols used in above equations. $\Gamma^{(\alpha)}(q)$, denotes the  baryon vertex dominated by the $1^+\,(0^+)$ diquark, and the two baryon vertexes are now coupled together for the existence of the kernels $K_{10}$ and $K_{01}$. Also notice there are four different interaction kernels, where $K_{11}^{\alpha\beta}(P,k,q)$ denotes the one responsible for the $1^+ \to 1^+$ heavy diquark transition, and then $K^\alpha_{01}$, $K^\beta_{10}$, and $K_{00}$ are implied. As usual, $M$ represents the mass of the baryon and we have $P^2=M^2$. The constituent mass of the diquark is now represented by $M_{1(i)} $ with $i=0$ for the $0^+$ and $i=1$ for the $1^+$ diquark. The inner relative momenta $q$  and $k$ are now defined as 
\[q=\alpha_{2(i)}p_1-\alpha_{1(i)}p_2, ~~k=\alpha_{2(i)}k_1-\alpha_{1(i)} k_2,\]
with $\alpha_{1(i)}\equiv \frac{M_{1(i)}}{M_{1(i)}+M_2}$ and $\alpha_{2(i)}=1-\alpha_{1(i)}$. $S(k_2)=i\frac{1}{\slashed k_2 - M_2}$ is the Dirac propagator of the third quark as usual, while the effective propagator of the $1^+$ diquark reads\,\cite{LiQ2020},
\begin{gather*} 
D^{\alpha\beta}(p_1)  =D(p_1)d^{\alpha\beta}(p_{1\perp}),~~~d^{\alpha\beta}(p_{1\perp})= -g^{\alpha\beta}+\frac{p_{1\perp}^\alpha p_{1\perp}^\beta}{M_{1(1)}^2},
\end{gather*}
where $D(p_1)=i\frac{1}{p_1^2-M_{1(1)}^2+i\epsilon}$ is the usual scalar propagator.

We define two Bethe-Salpeter wave functions as usual,
\begin{align}
  B_\alpha(P,q,\xi)& \equiv S(p_2) \Gamma^\beta(P,q,\xi) D_{\alpha\beta}(p_1) , \label{E-BSwave-1}\\
  B(P,q,\xi)&\equiv S(p_2) \Gamma(P,q,\xi) D(p_1). \label{E-BSwave-2}
% \varphi_\beta(P,q_\perp) \equiv - i\int \frac{\up d q_P}{2\pi} \psi_\beta(P,q).
\end{align}
Here the $B_\alpha(P,q,\xi)$ and $B(P,q,\xi)$ represent the wave functions of the baryons formed by the axialvector and scalar diquarks, respectively. The total momentum $P$ and the polarization state $\xi$ in the BS wave function $B_{(\alpha)}(P,q,\xi)$ and vertex $\Gamma^{(\beta)}(P,q,\xi)$ would be omitted unless it is necessary to specify them. The $\frac{1}{2}^+$ baryons with different flavors can be generally expressed as $\begin{pmatrix} B^\alpha \\ B \end{pmatrix}$, with $\begin{pmatrix} B^\alpha \\ 0 \end{pmatrix}$ and $\begin{pmatrix} 0 \\ B \end{pmatrix}$ describing the baryons with flavor-symmetric and antisymmetric $(bc)$-diquark inside respectively. Note now the mixing effects are naturally reflected in the wave functions. 
Using the wave functions, the coupled BSEs in \eref{E-BSE-A1} and \eref{E-BSE-A0} can also be rewritten in the matrix form 
\begin{gather}
S^{-1}(p_2) \begin{bmatrix} (D^{\alpha\beta})^{-1}(p_1) &0 \\ 0 &D^{-1}(p_1) \end{bmatrix} \begin{pmatrix} B_\beta[q] \\ B[q] \end{pmatrix}=  \int \frac{\d^4k}{(2\pi)^4}
(-i)\begin{bmatrix} K_{11}^{\alpha\beta} & K_{01}^{\alpha}\\ K_{10}^{\beta} &  K_{00} \end{bmatrix} 
\begin{pmatrix} B_\beta[k] \\ B[k] \end{pmatrix},
\end{gather} 
where $S^{-1}(p_2)$ and $[D^{(\alpha\beta)}]^{-1}(p_1)$ are the inverses of the corresponding propagators.
The normalization condition is generally expressed as,
\begin{align}
\int \frac{\d^4 q}{(2\pi)^4}\frac{\d^4 k}{(2\pi)^4}  \begin{pmatrix}\bar{B}_\alpha[q,\xi'] & \bar B[q,\xi']\end{pmatrix} \frac{\partial}{\partial P^0} \left[ \up{I}^{\alpha\beta}(P,k,q) \right]\begin{pmatrix}{B}_\beta[k, \xi] \\  B[k,\xi]\end{pmatrix}  = i2M \delta_{\xi\xi'},
\end{align}
where $\up{I}^{\alpha\beta}$ reads
\begin{align} \label{E-Mix-I}
\up{I}^{\alpha\beta}(P,k,q)&=(2\pi)^4 \delta^4(k-q) S^{-1}(p_2)\begin{bmatrix} [D^{\alpha\beta}(p_1)]^{-1} &0 \\ 0 &D^{-1}(p_1) \end{bmatrix}+ i\begin{bmatrix} K_{11}^{\alpha\beta} & K_{01}^{\alpha}\\ K_{10}^{\beta} &  K_{00} \end{bmatrix}.
\end{align}

\subsection{Salpeter equations under the instantaneous approximation}

Using the obtained form factors, the four interaction kernels can now be expressed as,
\begin{equation}
\left \{
\begin{aligned}
K^{\alpha\beta}_{11} (P,k,q) &= g^{\alpha\beta}(p_1+k_{1})^\nu [ C_{11} V(k-q)]\gamma_\nu ,\\
K^\alpha_{01}(P,k,q)&= i \frac{2w_{1(1)}}{M_{1(1)}M_{1(0)}}  \epsilon^{\alpha \nu  k_1 p_1} [ C_{01} V(k-q)]\gamma_\nu,\\
{K}_{00}(P,k,q) &=  (p_1+k_{1})^\nu [ C_{00}V(k-q)]\gamma_\nu,\\
K^\alpha_{10}(P,k,q)&= i  \frac{2w_{1(0)}}{M_{1(1)} M_{1(0)}}  \epsilon^{\alpha \nu k_1 p_1} [ C_{10}V(k-q)]\gamma_\nu,
\end{aligned}
\right.
\end{equation}
where $(p_1+k_1)=\varrho\hat P+(k+q)_\perp$ is dependent on both $M$ and $q_P$, and we used the abbreviation $\varrho=2(\alpha_1M+q_{P})$; the diquarks' kinetic energy parts $w_{1(i)}=(M_{1(i)}^2 - p_{1\perp}^2)^{\frac12}$ are different for the difference of the diquark mass $M_{1(1)}$ and $M_{1(0)}$. As usual, we deal with the BSEs under the instantaneous approximation, namely, $V(k-q)\sim V(k_\perp -q_\perp)$, and also we only consider the time component parts of the Lorentz structure in the kernels. Then we further define four instantaneous interaction kernels,
\begin{equation}
\left \{
\begin{aligned}
\varkappa_{11}^{\alpha\beta} (k_\perp,q_\perp)   &\equiv   \left[C_{11}V(k_\perp-q_\perp)\right] g^{\alpha\beta}  ,\\
\varkappa_{01}^{\alpha}(k_\perp,q_\perp)   &\equiv  [C_{01}V(k_\perp-q_\perp)]   i \frac{ \epsilon^{\alpha 0 k_{\!\perp} q_{\perp}} }{M_{1(0)}M_{1(1)}}  ,\\
\varkappa_{00}(k_\perp,q_\perp)  &\equiv   [C_{00} V(k_\perp-q_\perp) ]  , \\
\varkappa_{10}^{\alpha}(k_\perp,q_\perp)  &\equiv  [C_{10}V(k_\perp-q_\perp)]  i \frac{\epsilon^{\alpha 0 k_{\!\perp} q_{\perp}} }{M_{1(0)}M_{1(1)}},
\end{aligned}
\right.
\end{equation}
where the potential reads $V(k_\perp-q_\perp)=(2\pi)^3 \delta^3(k_\perp-q_\perp) V_1+V_2(k_\perp-q_\perp)$; notice that the $V_1$ part in the potential does not contribute to the $\varkappa_{01}^\alpha$ and the $\varkappa_{10}^\alpha$ since the Dirac delta function. The four instantaneous interaction kernels $\varkappa(k_\perp,q_\perp)$ here do not depend on the time component of $k$ or $q$, and the Lorentz structure factor $\gamma_0$ is also split out for later convenience. Now the BS vertexes can be expressed as
\begin{align}
\Gamma^\alpha(q) &= \gamma_0\left[ \varrho \Theta^\alpha_{11}(q_\perp) + 2w_{1(1)} \Theta_{01}^\alpha(q_\perp) \right], \label{E-Gamma1}\\
\Gamma(q) &= \gamma_0 \left [\varrho  \Theta_{00}(q_\perp) + 2w_{1(0)} \Theta_{01}(q_\perp) \right], \label{E-Gamma0}
\end{align}
where the 4 three-dimensional vertexes $\Theta_{11}^\alpha$, $\Theta_{01}^\alpha$, $\Theta_{00}$ and $\Theta_{10}$ are only explicitly dependent on $q_\perp$ and defined as,
\begin{equation}
\left \{
\begin{aligned}
\Theta_{11}^\alpha(q_\perp) &=\int \frac{\up d^3 k_\perp}{(2\pi)^3} \varkappa_{11}^{\alpha\beta}(k_\perp-q_\perp)  \varphi_\beta(k_\perp),\\
\Theta^\alpha_{01}(q_\perp) &=\int \frac{\up d^3 k_\perp}{(2\pi)^3} \varkappa^\alpha_{01}(k_\perp,q_\perp)  \varphi(k_\perp),\\
\Theta_{00}(q_\perp) &=\int \frac{\up d^3 k_\perp}{(2\pi)^3} \varkappa_{00}(k_\perp-q_\perp) \varphi(k_\perp),\\
\Theta_{10}(q_\perp) &=\int \frac{\up d^3 k_\perp}{(2\pi)^3} \varkappa^\alpha_{10}(k_\perp,q_\perp)  \varphi_\alpha(k_\perp).
\end{aligned}
\right.
\end{equation}

Performing the contour integral over $q_P$ on both sides of \eref{E-BSwave-1} and \eref{E-BSwave-2},  we can obtain the three-dimensional Salpeter equations (SE),
\begin{align}
\varphi_\alpha(q_\perp) &=d_{\alpha\beta}\left[\frac{\Lambda^+ \gamma_0 }{M-w_{1(1)}-w_2}  \left(\Theta^\beta_{11}+ \Theta^\beta_{01} \right) - \frac{\Lambda^- \gamma_0}{M+w_{1(1)}+w_2} \left( \Theta^\beta_{11} - \Theta^\beta_{01}  \right)\right], \\
\varphi(q_\perp) &=\frac{\Lambda^+ \gamma_0}{M-w_{1(0)}-w_2} \left(\Theta_{00}+ \Theta_{10} \right) - \frac{\Lambda^- \gamma_0}{M+w_{1(0)}+w_2} \left( \Theta_{00} - \Theta_{10}  \right),
\end{align}
where the projector operators are defined as $\Lambda^\pm(p_{2\perp})\gamma_0=\frac{1}{2}\left(1 \pm {H_2} \right)$, and the abbreviation $
H_2 \equiv \frac{1}{w_2}\left(\slashed p_{2\perp}+M_2\right)\gamma^0
$ is the usual Dirac Hamiltonian divided by $w_2$. Define the positive and negative energy wave functions as $\varphi^{\pm}_{(\alpha)}(q_\perp)=\Lambda^\pm \gamma^0 \varphi_{(\alpha)}$, and $\varphi_{(\alpha)}=\varphi_{(\alpha)}^++\varphi_{(\alpha)}^-$. Notice in the weak binding condition $M\sim (w_{1(i)}+w_2)$, $\varphi^+\gg \varphi^-$, and the positive energy wave function $\varphi^+(q_\perp)$ usually dominates in the Salpeter wave function $\varphi_{(\alpha)}$.  The SEs can be further rewritten as the following simple Shr\"odinger type
\begin{equation}\label{E-VS-SE10} 
\begin{aligned} 
M \varphi_\alpha &= \left (w_{1(1)}+w_2 \right) H_2\varphi_\alpha +   \left( H_2  \Theta_{11}^\beta  +  \Theta_{01}^\beta \right) d_{\alpha\beta} , \\
M \varphi &= \left (w_{1(0)}+w_2 \right) H_2\varphi +  \left( H_2   \Theta_{00} +  \Theta_{10}\right). 
\end{aligned}
\end{equation}
The obtained coupled SEs can also be written as the matrix form,
\begin{gather} \label{E-VS-SE}
M\begin{pmatrix}\varphi_\alpha \\ \varphi\end{pmatrix}=\left[\begin{pmatrix}w_{1(1)} & \\ & w_{1(0)} \end{pmatrix} +w_2 \right] H_2  \begin{pmatrix}\varphi_\alpha \\ \varphi\end{pmatrix} +\int \frac{\up d^3 k_\perp}{(2\pi)^3}  \begin{pmatrix} d_{\alpha\beta}  H_2 \varkappa_{11}^{\beta\gamma} &  d_{\alpha\beta} \varkappa_{01}^{\beta} \\
\varkappa_{10}^\gamma &  H_2 \varkappa_{00} \end{pmatrix} \begin{pmatrix}\varphi_\gamma \\ \varphi\end{pmatrix},
\end{gather}
which is the fundamental equation of this work.

%The spinor can be split out from the Salpeter wave function, $\begin{pmatrix}\varphi_\alpha \\ \varphi\end{pmatrix} = \begin{pmatrix}A_\alpha \\ A\end{pmatrix} u(P,r)$. 

The obtained three-dimensional BSE, namely, \eref{E-VS-SE}, indicates that the mass contributions of the $J^P=\frac{1}{2}^+$ baryons originate from two parts, the kinetic energy and the potential energy; in the later one, the mixing effects are included naturally in the Salpeter equation. Also notice \eref{E-VS-SE} is in fact the eigenvalue equation of the  wave function $\begin{pmatrix}\varphi_\alpha \\ \varphi\end{pmatrix}$ with the baryon mass $M$ as the eigenvalue.

The normalization of the Salpeter wave functions are similar as before.
Since we only consider the time component of the interaction kernel, $K^\alpha_{01}(P,k,q)\simeq K_{01}^{\alpha}(k_\perp,q_\perp)$ has no dependence on $P^0$ and $q_P$, and the anti-diagonal parts of the normalization kernel $\up{I}^{\alpha\beta}(P,k,q)$ in \eref{E-Mix-I} vanish after the partial differential over $P^0$. The inverse of the vector propagator reads,
\[ D^{-1}_{\alpha\beta}(p_1)=\vartheta_{\alpha\beta}D^{-1}(p_1),\]
where $\vartheta^{\alpha\beta}\equiv - (g^{\alpha\beta}+{p^\alpha_{1\perp} p^\beta_{1\perp}}/{w_{1(1)}^2} )$ and  fulfills $\vartheta^{\alpha\beta}d_{\beta\gamma}=\delta^{\alpha}_{\gamma}$; and now  there is no $P^0$ dependence in the numerator of $D^{-1}_{\alpha\beta}(p_1)$. 
Inserting the inverses of the propagators, we further obtain the normalization condition for the general Salpeter wave function,
\begin{equation}\label{E-Norm-D1}
\int \frac{\d^3 q_\perp}{(2\pi)^3}  %\vartheta^{\alpha\beta}[ \bar\varphi_\alpha(\bar r) \Delta h_1  \varphi_\beta(r)] +\bar\varphi(\bar r) \Delta h_0 \varphi(r)
\frac1M \begin{pmatrix}\bar \varphi_\alpha [\xi'] & \bar\varphi[\xi']  \end{pmatrix} \begin{bmatrix} \vartheta^{\alpha\beta}  O_{1} &  0 \\ 0  &O_{0}  \end{bmatrix}  \begin{pmatrix} \varphi_\beta[\xi] \\ \varphi[\xi] \end{pmatrix}
=\delta_{\xi\xi'},
\end{equation}
with the operator $O_{i}\equiv  \gamma^0\left[ \alpha_{1(i)} M H_2+ w_{q(i)} \right]$, and $w_{q(i)}  \equiv \alpha_{2(i)} w_{1(i)}-\alpha_{1(i)} w_2$, where $i=1(0) $ denotes the $1^+(0^+)$ heavy diquark.

\subsection{Salpeter wave functions for the baryons with different flavors}

The $J^P=\frac{1}{2}^+$ Salpeter wave function formed by the $1^+$ diquark and a $\frac{1}{2}^+$ light quark has been given in the previous work\,\cite{LiQ2020},
\begin{gather} \label{E-wave-1-2Z}
\varphi_{\alpha}(P,q_\perp,\xi) =  A_\alpha(q_\perp)\gamma^5 u(P,r) ,
\end{gather}
with
\[
A_\alpha= \left( g_1+ g_2 \slashed y  \right)(\gamma_\alpha + \hat P_\alpha)  + \left( g_3+ g_4 \slashed y \right) y_{\alpha},
\]
where the radial wave function $g_i(\vabs{q})~(i=1,\cdots,4)$ just depends on $\vabs{q}$ explicitly. It is clear to see that $g_1$ corresponds to the $S$-wave component, $g_{2(3)}$ the $P$-wave components, and $g_4$ contributes to both the $S$ and $D$ partial waves (see Ref.\,\cite{LiQ2020} for a detailed expression in terms of the spherical harmonics $Y_l^m$). %Notice the wave function of $J^P=\frac{1}{2}^+$ baryon formed by a $1^+$ diquark and a third quark\,\cite{LiQ2020},  

Combined with the $\frac{1}{2}^+$ Salpeter wave function formed by the $0^+$ diquark and a $\frac12^+$ quark, namely, 
\begin{gather}\label{E-wave-D0-1H}
\varphi(P,q_\perp,\xi)=A(q_\perp) u(P,r),~~~A(q_\perp)\equiv g_5 + g_6 \slashed y,
\end{gather} 
we construct the Salpeter wave function for the $J^P=\frac{1}{2}^+$ baryons with different flavors as
\begin{gather}\label{E-wave}
\begin{pmatrix} \varphi_\alpha (P,q_\perp,\xi) \\ \varphi(P,q_\perp,\xi) \end{pmatrix} =
\begin{pmatrix} A_\alpha (q_\perp)  \\ A(q_\perp) \end{pmatrix} u(P,\xi).
\end{gather}
By inserting the wave functions into \eref{E-Norm-D1} and summing over the spin states, we obtain the following normalization condition,
\begin{equation}\label{E-Norm-Mix}
\int \frac{\d^3 q_\perp}{(2\pi)^3} \frac{1}{2M}\up{tr}\left(\vartheta^{\alpha\beta} \bar A_\alpha  O_{1} A_\beta + \bar A O_{0} A \right)({\slashed P}+M)=1.
\end{equation}
From this normalization, we can define two integrals $\up{I}_{(1)}$ and $\up{I}_{(0)}$ to represent the percentages of the the two components in the baryon Salpeter wave function,
\begin{equation}\label{E-Norm-I10}
\begin{aligned}
\up{I}_{1} &\equiv \int \frac{\d^3 q_\perp}{(2\pi)^3} \frac{1}{2M}\up{tr}\left(\vartheta^{\alpha\beta} \bar A_\alpha O_{1}  A_\beta \right)({\slashed P}+M),\\
\up{I}_{0} &\equiv \int \frac{\d^3 q_\perp}{(2\pi)^3} \frac{1}{2M}\up{tr}\left(\bar A O_{0}  A \right)({\slashed P}+M),
\end{aligned}
\end{equation}
where we have $\up{I}_{1}+\up{I}_{0}=1$; it can be seen obviously that $\up{I}_{1}\,(\up{I}_{0})$ denotes the weight of a baryon with its $(bc)$-diquark core in $1^+\,(0^+)$ state, or equivalently speaking, in the flavor-symmetric\,(antisymmetric) state. Namely, the methods developed here also allow us to investigate the mixing effects of in the $\frac12^+$ $(bcq)$ baryons, and the numerical results will presented in the following section.

\section{Mass spectra and wave functions of the $(bcq)$ baryons}\label{Sec-5}
For the $J^P=\frac{1}{2}^+$ $(bcu)$ and $(bcs)$ baryons with pure $0^+$ $(cb)$-diquarks, the parameters $V_0$ used are $-0.360$ and $-0.336$\,GeV respectively, which are determined by the spin-weighted average methods\,\cite{LiQ2020} by fitting to the corresponding meson spectra. 
\begin{table}[h!]
\caption{Mass spectra\,(in GeV) of the low lying $J^P=\frac{1}{2}^+$ $(bcq)$ baryons.  The baryon states are labeled by the notations $N\!L(n^{2S+1}L_J)$, where $N\!L$ describe the radial and orbital quantum numbers between the $(bc)$-diquarks and the light quark, while $(n^{2S+1}L_J)$ are used to describe the corresponding  states of the inside $(bc)$-diquarks. Type I represents the results obtained from the pure $1^+$ and $0^+$ $(bc)$-diquarks, while type II contains the mixing effects between the $1^+$ and $0^+$ $(bc)$-diquarks insider the baryons obtained by solving \eref{E-VS-SE}.}\label{Tab-Mass-D0}
\vspace{0.2em}\centering
\setlength{\tabcolsep}{3pt} % \renewcommand{\arraystretch}{1.5}
\begin{tabular}{  c|ccccc }
\toprule[1.5pt]
%$N^{2S+1}L_J(nl)$ &$(bcu)$   &$(bcd)$  &$(bcs)$\\
%\midrule[1.2pt]
%$1^1S_{1/2}(1S)$   &$6.948$	 &$6.950$  &$7.051$ \\
%$1^1S_{1/2}(2S)$   &$7.430$  &$7.433$  &$7.547$ \\
%$2^1S_{1/2}(1S)$   &$7.465$  &$7.468$  &$7.579$ \\
Baryons &type    &$1S(1^3S_1), 1S(1^1S_0)$ 	&$1S(2^3S_1),1S(2^1S_0)$		&$[2S(1^3S_1),1D(1^3S_1)],2S(1^1S_0)$ \\
\midrule[1.1pt] 
\multirow{2}{*}{$(M_{\Xi^+_{bc}},M_{\Xi'^+_{bc}})$}&I            & (6.931, 6.948)     &(7.417, 7.430)  & ([7.446, 7.463], 7.465) \\  
&II           & (6.930, 6.942)     &(7.401, 7.413)  & ([7.444, 7.463], 7.458) \\  
\midrule[1.1pt] 
\multirow{2}{*}{$(M_{\Xi^0_{bc}},M_{\Xi'^0_{bc}})$}&I            & (6.934, 6.950)     &(7.420, 7.433)  & ([7.450, 7.467], 7.468) \\  
&II           & (6.933, 6.945)     &(7.404, 7.416)  & ([7.448, 7.467], 7.461) \\  
\midrule[1.1pt]
\multirow{2}{*}{$(M_{\Omega_{bc}},M_{\Omega^\prime_{bc}})$}&I            & (7.033, 7.051)     &(7.531, 7.547)  & ([7.560, 7.598], 7.579) \\  
&II           & (7.032, 7.045)     &(7.516, 7.529)  & ([7.558, 7.598], 7.572) \\  
\bottomrule[1.5pt]
\end{tabular}
\end{table}

\begin{figure}[h!]
\centering
\includegraphics[width = 0.6\textwidth, angle=0]{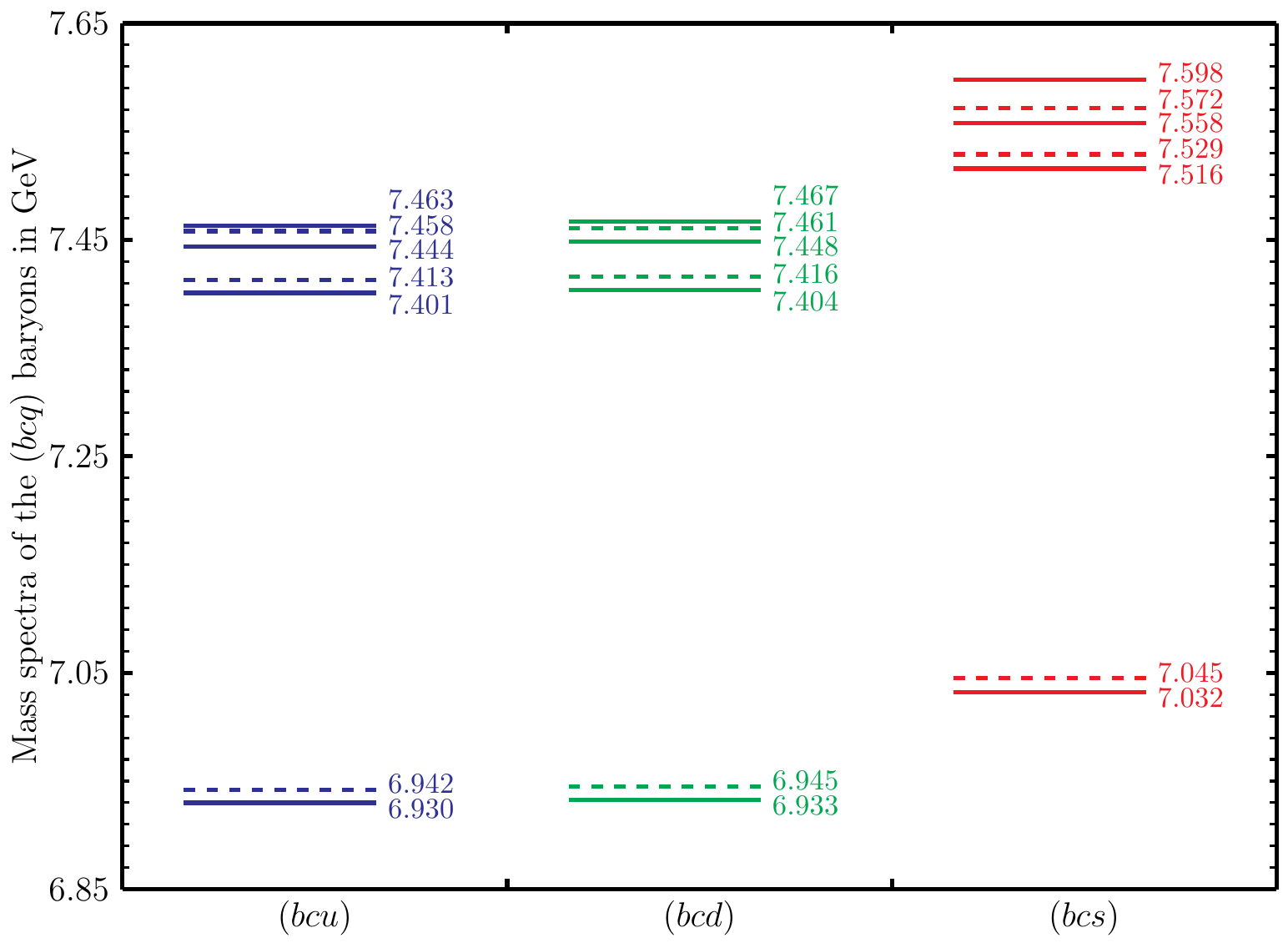}
\caption{Mass spectra\,(in GeV) of the low-lying $J^P=\frac{1}{2}^+$ $(bcq)$ baryons. The solid\,(dotted) line denotes the baryons with the inside $(bc)$-diquarks in the flavor-symmetric\,(anti-symmetric). }\label{Fig-bcq-masses}
\end{figure}
The obtained mass spectra for the $\frac{1}{2}^+$ $(bcq)$ baryons are listed in \autoref{Tab-Mass-D0},  with both the $1S$ and $2S$ $(bc)$-diquark cores. Type I are the results from the pure $1^+$ and $0^+$ $(bc)$-diquarks, while type II are obtained by solving \eref{E-VS-SE} which has incorporated the mixing effects between the $1^+$ and $0^+$ $(bc)$-diquarks insider the baryons. Here, to reflect the dominant characteristic, we still label baryon states by the nonrelativistic notations $N\!L(n^{2S+1}L_J)$ with the first two quantum numbers $N\!L$ describing the baryons formed by the diquarks and third light quark, while $n^{2S+1}L_J$ describing the states of diquarks; for example, $1S(2^{3}S_1)$ denotes the radial quantum number of the $(bc)$-diquark core is $n=2$ with its orbital angular momentum being in $S$-wave, and the radial quantum number of the baryon is $N=1$ with the orbital angular momentum of the diquark-quark being in $S$-wave. Our results show that the masses of the $2S(1^3S_1)$, $1D(1^3S_1)$ $\Xi_{bc}$ and the $2S(1^1S_0)$ $\Xi'_{bc}$ are very close, and theses three states locate in the mass range $7.440\sim 7.470\,\si{GeV}$. 

\begin{figure}[h!]
\centering
\subfigure[$\Xi_{bc}$ in $1S(1^3S_1)$]   {\includegraphics[width=0.48\textwidth]{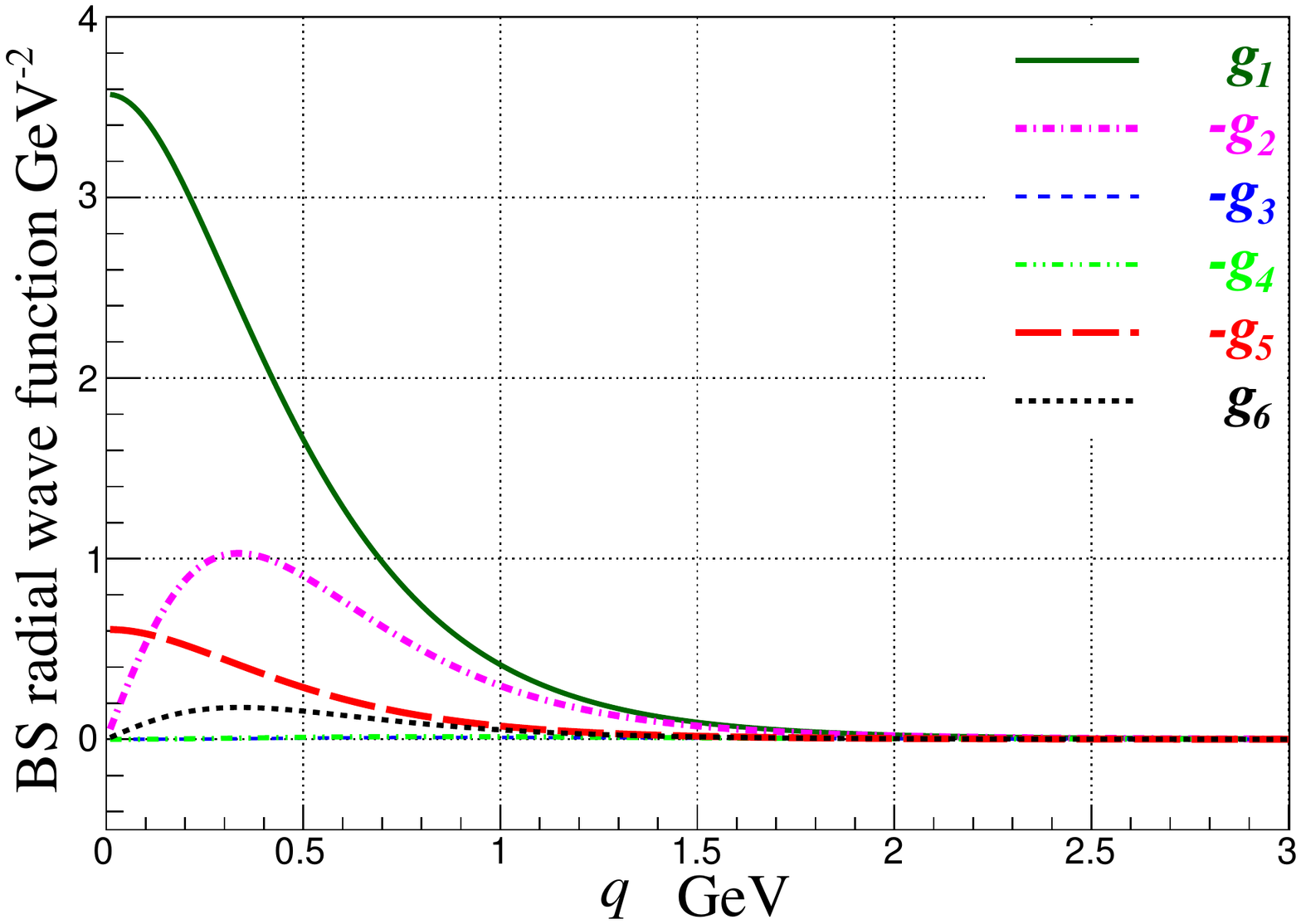} \label{Fig-Xibcu-wave-n1}}
\subfigure[$\Xi'_{bc}$ in $1S(1^1S_0)$]   {\includegraphics[width=0.48\textwidth]{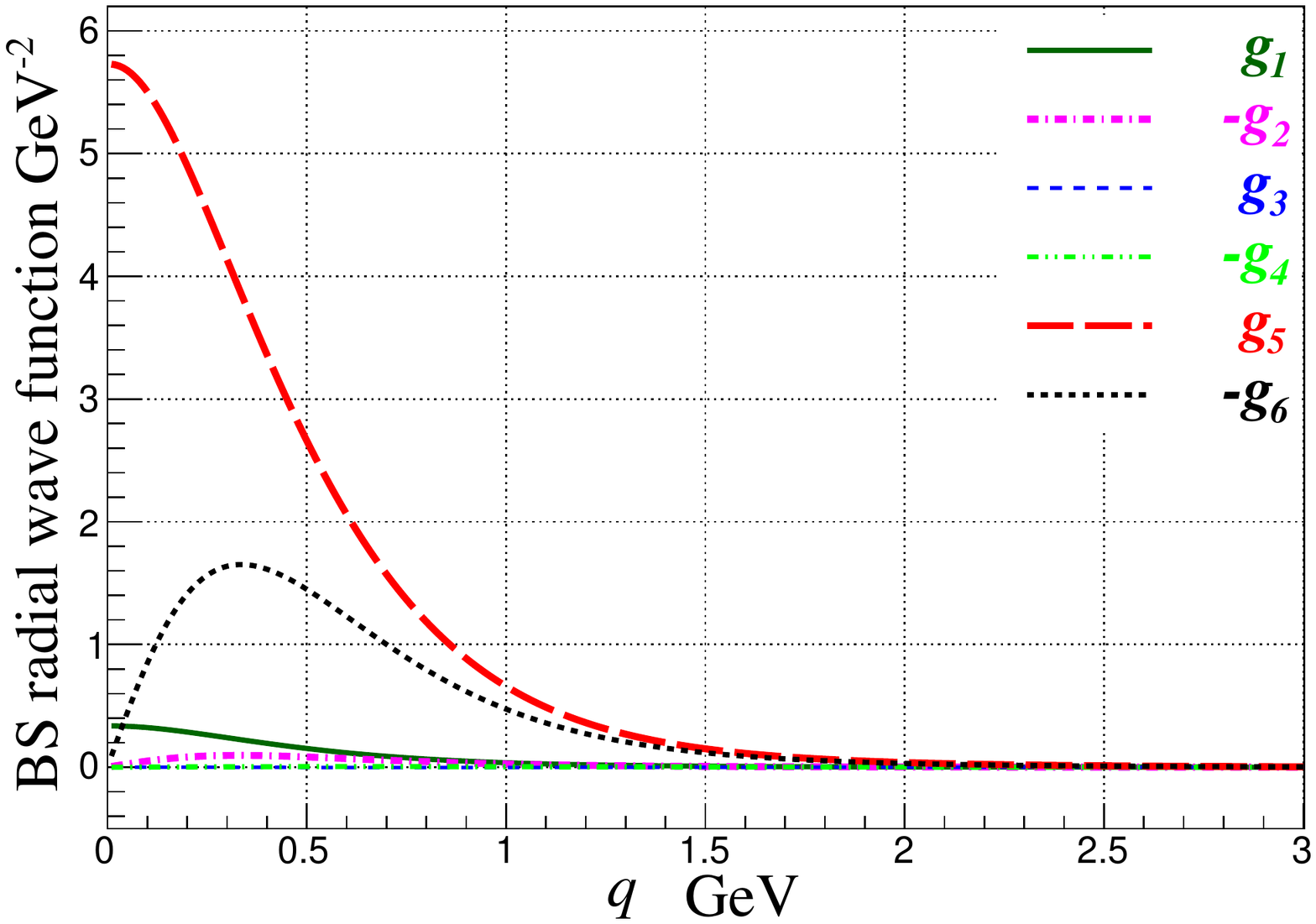} \label{Fig-Xibcu-wave-n2}}
\subfigure[$\Xi_{bc}$ in $2S(1^3S_1)$]   {\includegraphics[width=0.48\textwidth]{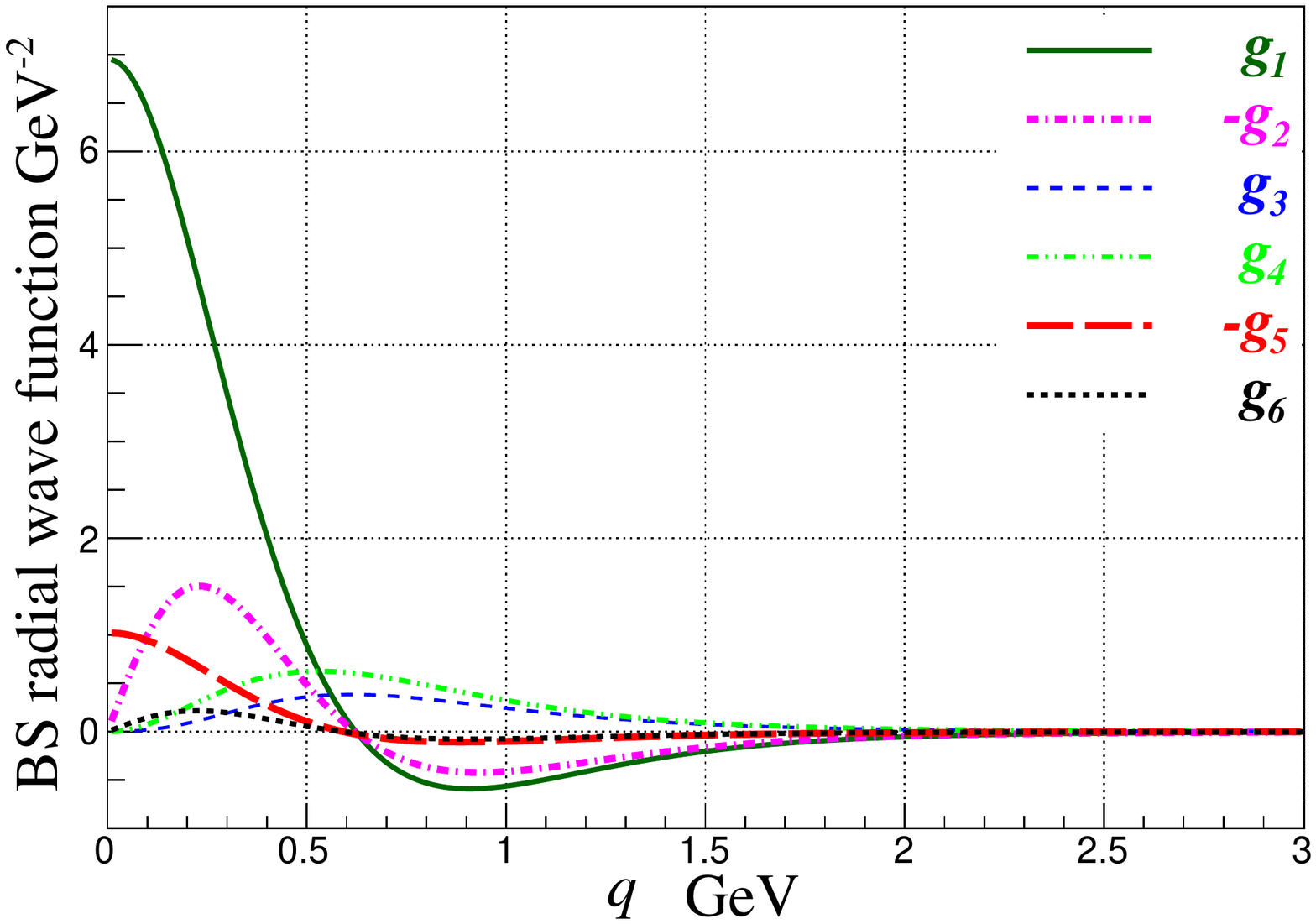} \label{Fig-Xibcu-wave-n3}}
\subfigure[$\Xi'_{bc}$ in $2S(1^1S_0)$]   {\includegraphics[width=0.48\textwidth]{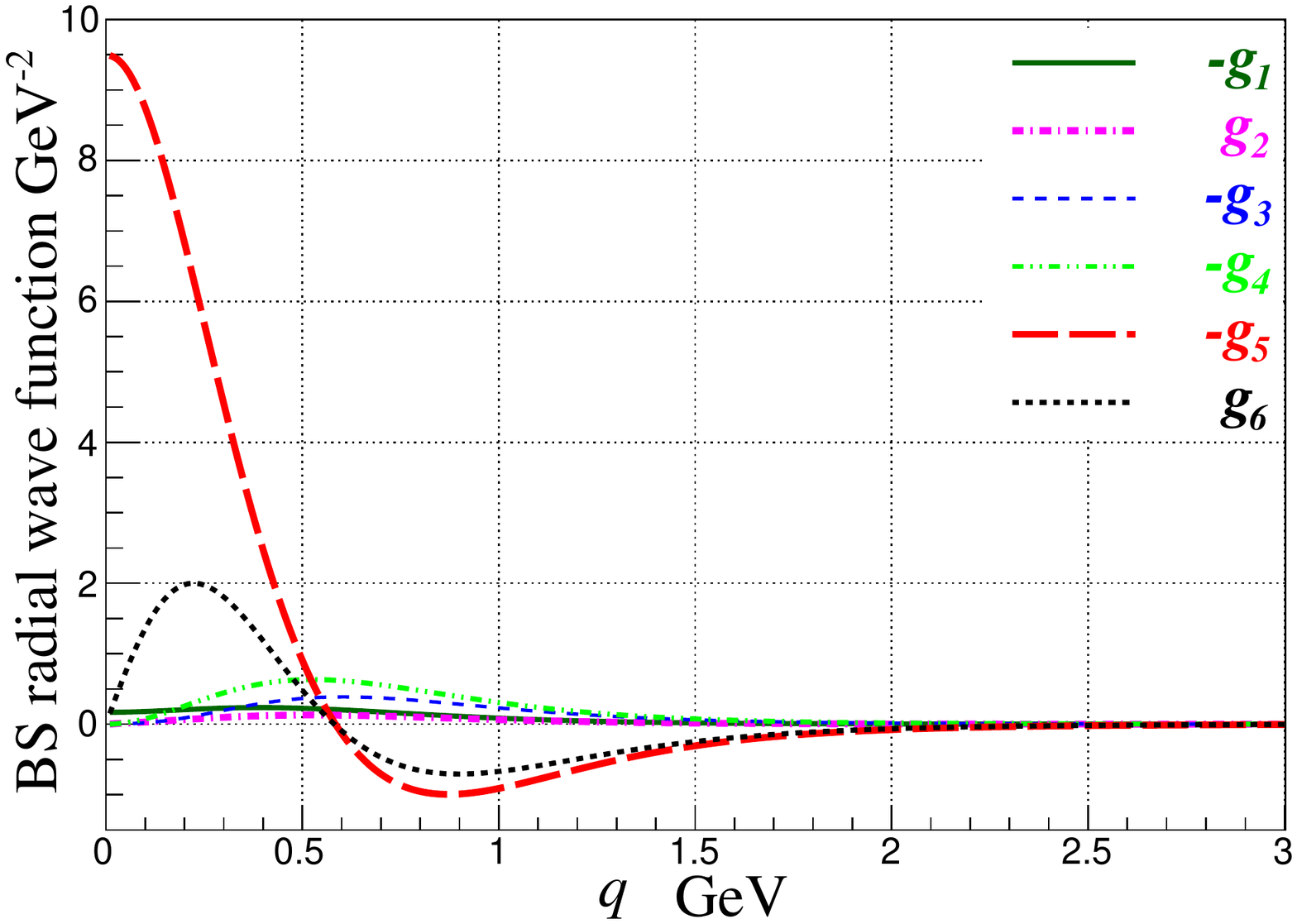} \label{Fig-Xibcu-wave-n4}}
\subfigure[$\Xi_{bc}$ in $1D(1^3S_1)$]   {\includegraphics[width=0.48\textwidth]{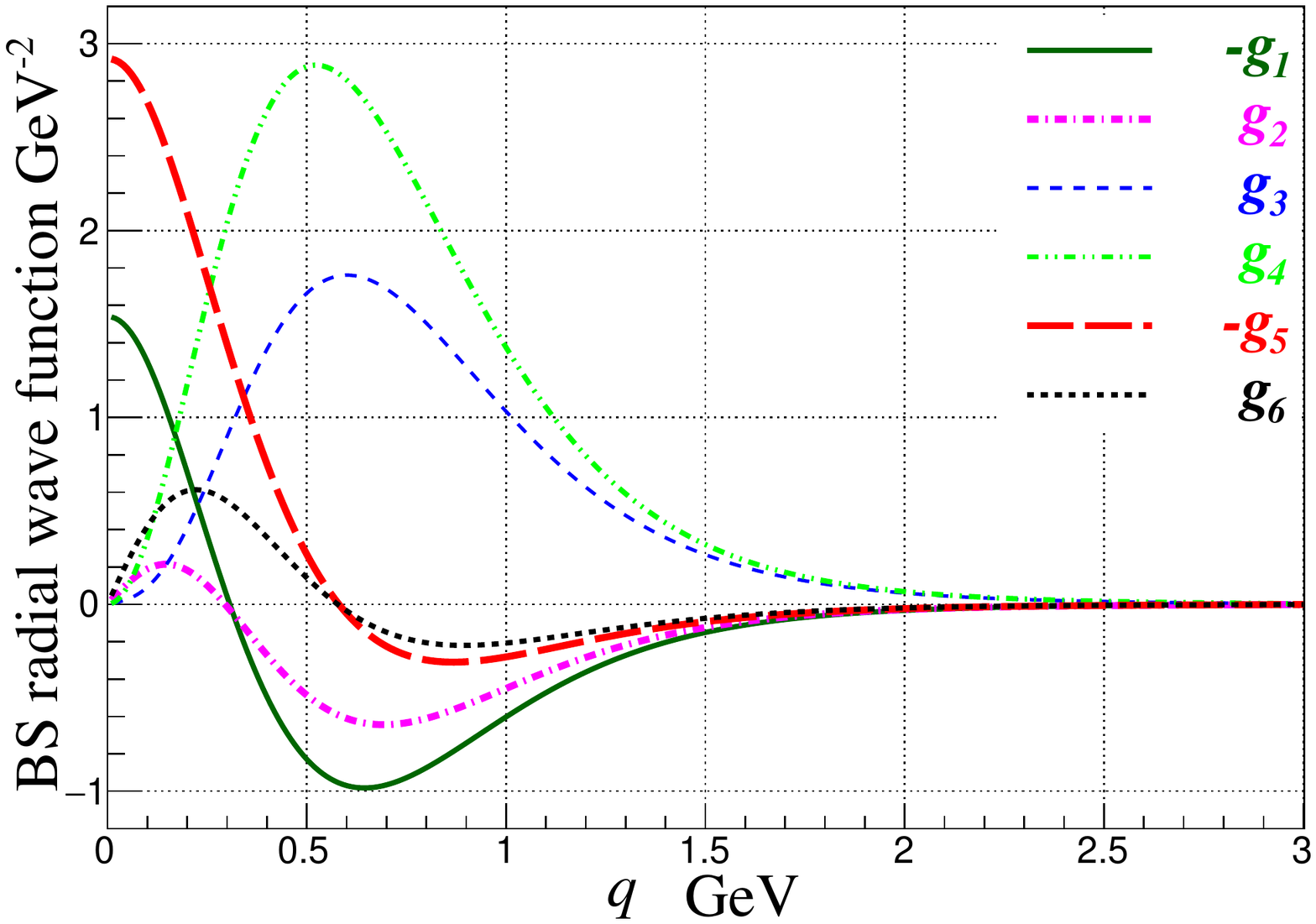} \label{Fig-Xibcu-wave-n5}}
\caption{BS radial wave functions of the $J^P=\frac{1}{2}^+$ $(bcu)$ baryons for the first five energy levels with the $S$-wave $(bc)$-diquark cores. } \label{Fig-Xibcu-wave}
\end{figure}

\autoref{Fig-Xibcu-wave} graphically shows the BS wave functions of the $\frac{1}{2}^+$ $(bcu)$ baryons in the first five energy levels when the $(bc)$-diquarks are in the $S$-wave states. Note that $g_{1-4}$ and $g_{5-6}$ correspond to the baryons formed by the $1^+$ and $0^+$ $(bc)$-diquark cores, respectively. Firstly note that in the relativistic wave functions, $\frac{1}{2}^+$ $(bcq)$ baryons contain all the $S$-, $P$-, and $D$-wave components. By analyzing the node structures and wave forms, we can find that, though $\Xi_{bc}$ and $\Xi'_{bc}$ are dominated by the states with $1^+$ and $0^+$ $(bc)$-diquark cores respectively, there still exist the mixing effects from each other. By calculating the $\up I_1$ and $\up I_0$ in \eref{E-Norm-I10}, we obtain that $\up{I}_1(\Xi_{bc})$ for the $1S(1^3S_1)$ and $2S(1^3S_1)$ states are both about $98.8\%$, namely, the mixing components from the  baryon states with $0^+$ diquark are $\sim\!1.2\%$. However, for $\Xi_{bc}$ in the $1D(1^3S_1)$ state, the mixing effects from the $\Xi'_{bc}[2S(1^1S_0)]$ can reach $\sim\!9.2\%$ and hence cannot be ignored, which can also be seen from \autoref{Fig-Xibcu-wave-n5} directly. For $\Omega_{bc}^{(\prime)}$, the obtained mixing effect are less than $1\%$ in all the low lying states.
The numerical results indicate the mixing effects  are quite small between the flavor-symmetric and -antisymmetric $(bc)$-diquarks in $\Xi^{(\prime)}_{bc}$ and $\Omega^{(\prime)}_{bc}$ in the ground and the first radially excited states. The small mixing also states that it is a good choice to take $(bc)$ as the diquark with definite spin in the ground states of $\Xi^{(\prime)}_{bc}$ and $\Omega^{(\prime)}_{bc}$ systems. Besides the mixing effects from the $\Xi_{bc}'[2S(1^1S_0)]$ state, \autoref{Fig-Xibcu-wave-n5} also shows that both the $2S$\,(dark green solid) and $1D$\,(light green dotted) partial waves have great contributions to the $1D(1^3S_1)$ state and the $2S$-$1D$ mixing effects should not be omitted either in analysis.

%---------------------------------------------------------------------------
\begin{table}[h!]
\caption{Comparisons of the mass predictions for the $\frac{1}{2}^+$ $(bcq)$ baryons (in GeV).  The unprimed $\Xi_{bc}$ and $\Omega_{bc}$ denote the baryons with $1^+$ $(bc)$-diquark cores, and the primed ones correspond to those with $0^+$ $(bc)$-diquark cores. }\label{Tab-Mass-C}
\vspace{0.2em}\centering
\begin{tabular}{ cccccccccccc }
\toprule[2pt]
Baryon 			& This	 &\cite{Brown2014}&\cite{Karliner2014,Karliner2018}&\cite{Ebert2002}	&\cite{Albertus2007}	 &\cite{Giannuzzi2009} 	&\cite{WengXZ2018} &\cite{Ebert1997} &\cite{Roncaglia1995}\\
\midrule[1.5pt]
$\Xi_{bc}$		&6.930 &6.959				&6.914 			&6.933			&6.919				&6.904			&6.922			&6.950		&6.990\\
$\Xi'_{bc}$		&6.942 &6.943				&6.933 			&6.963			&6.948				&6.920			&6.948			&7.000		&7.040\\
$\Omega_{bc}$	&7.032 &7.032				&7.013 			&7.088			&6.986				&7.136			&7.011			&7.050		&7.060\\
$\Omega'_{bc}$	&7.045 &6.998				&7.025 			&7.116			&7.009				&7.165			&7.047			&7.090		&7.090\\
\bottomrule[2pt]
\end{tabular}
\end{table}

%---------------------------------------------------------------------------
\begin{table}[h!]
\caption{Comparisons of the mass splittings $(M_{\Xi_{bc}'}-M_{\Xi_{bc}})$ and $(M_{\Omega_{bc}'}-M_{\Omega_{bc}})$ in MeV.  The unprimed $\Xi_{bc}$ and $\Omega_{bc}$ denote the $\frac{1}{2}^+$ $(bcq)$ baryons with $1^+$ $(bc)$-diquark corse, and the primed ones correspond to those with $0^+$ $(bc)$-diquark cores. }\label{Tab-Mass-C2}
\vspace{0.2em}\centering
\begin{tabular}{ cccccccccccc }
\toprule[2pt]
Baryon 									& This	 			&\cite{Brown2014}				&\cite{Karliner2014,Karliner2018}&\cite{Ebert2002}	&\cite{Albertus2007}	 &\cite{Giannuzzi2009} 	&\cite{WengXZ2018} &\cite{Ebert1997} &\cite{Roncaglia1995}\\
\midrule[1.5pt]
$\left(M_{\Xi_{bc}'}-M_{\Xi_{bc}}\right)$			&12 &-16			&19 				&30				&29				&16					&26			&5		&50\\
$\left(M_{\Omega_{bc}'}-M_{\Omega_{bc}}\right)$	&13 &-34		&12 				&28				&23				&29					&36			&40	&30\\
\bottomrule[2pt]
\end{tabular}
\end{table}

\autoref{Tab-Mass-C} shows a comparison of our predictions for the ground state baryons with others'. Our results are generally consistent with others'. The mass of $\Omega_{bc}$ is about 100\,MeV greater than the corresponding $\Xi_{bc}$, which is consistent with the relevant meson spectra, $M_{D^{(*)}_s}-M_{D^{(*)0}}=104\,(105)\,\si{MeV}$, and $M_{B^{(*)}_s}-M_{B^{(*)}}=88\,(90)\,\si{MeV}$. The mass splitting between $\Xi_{bc}'(\Omega_{bc}')$ and $\Xi_{bc}(\Omega_{bc})$ is another interesting point and the corresponding comparisons are listed in \autoref{Tab-Mass-C2}, in which the average values for $(M_{\Xi_{bc}'}-M_{\Xi_{bc}})$ and $(M_{\Omega_{bc}'}-M_{\Omega_{bc}})$ are about 19 and 20\,MeV respectively. Notice that the lattice results in Ref.\,\cite{Brown2014} show the different signs with others'; also in Refs.\,\cite{Roncaglia1995,Karliner2014,Karliner2018,WengXZ2018} the $\frac{1}{2}^+$ $(bcq)$ baryon doublet is labeled by the definite spin of the $(cq)$-diquark which is different from the definite spin of the $(bc)$-diquark used here. By using the Clebsch-Gordan coefficients, the baryon under different diquark basis can be expressed as
\begin{equation}
\begin{bmatrix} \ket{(bc)_0q} \\ \ket{(bc)_1q} \end{bmatrix}=
\begin{bmatrix} -\frac{1}{2}& -\frac{\sqrt{3}}{2} \\ +\frac{\sqrt{3}}{2} & -\frac{1}{2}\end{bmatrix}
\begin{bmatrix} \ket{b(cq)_0} \\ \ket{b(cq)_1} \end{bmatrix}=
%\begin{bmatrix} \cos\theta & -\sin\theta \\ \sin\theta & \cos\theta\end{bmatrix}
%\begin{bmatrix} \ket{1(23)_0} \\ \ket{1(23)_1} \end{bmatrix},~\theta=+120^\circ;\\
%\begin{bmatrix} \ket{(12)_03} \\ \ket{(12)_13} \end{bmatrix}=
\begin{bmatrix} -\frac{1}{2}& +\frac{\sqrt{3}}{2} \\ -\frac{\sqrt{3}}{2} & -\frac{1}{2}\end{bmatrix}
\begin{bmatrix} \ket{(qb)_0c} \\ \ket{(qb)_1c} \end{bmatrix},%=
%\begin{bmatrix} \cos\theta & -\sin\theta \\ \sin\theta & \cos\theta\end{bmatrix}
%\begin{bmatrix} \ket{(31)_02} \\ \ket{(31)_12} \end{bmatrix},~\theta=-120^\circ.
\end{equation}
where $\ket{(bc)_0q}$ denotes the baryon state when the $b$- and $c$-quark inside the baryon form the spin-0 diquark, and then others are implied. Above relations can be considered a rotation within different diquark basis and the rotation angles are respectively $120^\circ$ and $-120^\circ$.

\section{Summary}\label{Sec-6}
In this work, based on the diquark picture and the instantaneous approximation, we have built the theoretical framework to deal with the doubly heavy baryons with $(bcq)$ flavors, especially the mixing effects between the $1^+$ and $0^+$ $(bc)$-diquark cores inside the baryons are considered in the Bethe-Salpeter kernel and wave functions. We constructed the Salpeter wave functions for the $J^P=\frac{1}{2}^+$ doubly heavy baryons. The mass spectra $M[{\Xi^{(')}_{bc}}]=6.930\,(6.942)\,\si{GeV}$ and $M[\Omega^{(')}_{bc}]=7.032\,(7.045)\,\si{GeV}$ are obtained by solving the corresponding BSE. Our predictions are generally consistent with other researches. By using the obtained BS wave functions, we calculated the mixing effects in $\Xi_{bc}-\Xi'_{bc}$ and $\Omega_{bc}-\Omega'_{bc}$, and the numerical results show that there only exist quite small\,($\sim\!1\%$) mixing effects between the $\frac{1}{2}^+$ $(bcq)$ baryons in ground states; while the mixing effects can reach $\sim\!10\%$ in the $1D$ $\Xi_{bc}$ and $\Xi_{bc}'$.  The obtained BS wave function would make it possible to do the precise calculations on the lifetimes, production and decays of the corresponding $\Xi^{(\prime)}_{bc}$ and  $\Omega^{(\prime)}_{bc}$.

\appendix
\section{Some expressions and derivations}\label{App-1}
%\subsection{\textup{The form factor of $0^+\to 1^+$ $(bc)$ transition}} \label{A-1}
%For the case of $0^+\to 1^+$ diquark transition, the corresponding form factor can be expressed as
%\begin{align*}
%\Sigma^{\alpha\mu}_1(0^+\to 1^+)
%&=-\int \frac{\up{D}^4 s}{(2\pi)^4} \frac{\up{D}^4 s'}{(2\pi)^4}(2\pi)^4\delta^4(s_1-s_1')  \up{Tr}~\bar{\Gamma}^\alpha_c(p',s') S(s_1)\Gamma_c(p,s) S(-s_2) \gamma^\mu S(-s'_2).
%\end{align*}
%Since $\Sigma^{\alpha\mu}_1(0^+\to 1^+)=-[\Sigma^{\alpha\mu}_1(0^+\to 1^+)]^\dagger$, then 
%\begin{align*}
%\Sigma^{\alpha\mu}_1(0^+\to 1^+)
%&=\int \frac{\up{D}^4 s}{(2\pi)^4} \frac{\up{D}^4 s'}{(2\pi)^4}(2\pi)^4\delta^4(s_1-s_1')  \up{Tr}~S^\dagger(-s'_2 ) (\gamma^\mu)^\dagger S^\dagger(-s_2) \Gamma^\dagger_c(p,s)  S^\dagger(s_1) [ \bar{\Gamma}^\alpha_c(p',s')]^\dagger \\
%&=\int \frac{\up{D}^4 s}{(2\pi)^4} \frac{\up{D}^4 s'}{(2\pi)^4}(2\pi)^4\delta^4(s_1-s_1')  \up{Tr}~\bar \Gamma_c(p,s)  S(s'_1) \Gamma^\alpha_c(p',s') S(-s'_2 )\gamma^\mu S(-s_2)  \\
%&=\int \frac{\up{D}^4 s'}{(2\pi)^4} \up{Tr}~\bar \Gamma_c(p,s)  S(s_1) \Gamma^\alpha_c(p',s') S(-s'_2 )\gamma^\mu S(-s_2)
%%&=\int \frac{\up{D}^4 s'}{(2\pi)^4} \up{Tr}~\bar \psi_c(p,s)  \Gamma^\alpha_c(p',s') S(-s'_2 )\\
%\end{align*}
%Comparing this result with the expression in \eref{E-formD1D0-4D}, we can achieve that 
%\begin{gather*}
%\Sigma^{\alpha\mu}_1(0^+\to 1^+)=-i\rho_1 \epsilon^{\alpha\mu p'p} = i\rho_1 \epsilon^{\alpha\mu pp'}=\Sigma^{\alpha\mu}_1(1^+\to 0^+),
%\end{gather*}
%which indicates that $\Sigma^{\alpha\mu}(0^+\to 1^+)=\Sigma^{\alpha\mu}(1^+\to 0^+)$.

\subsection{\textup{Salpeter equation with $0^+$ diquark core}} \label{A-Salpeter}
In the instantaneous approximation, the BS wave function of the baryon with $0^+$ diquark core can be expressed by the three-dimensional vertex as
\begin{gather}\label{E-BS-BD0-1}
B(q)=S(p_2) [\gamma^0  \Theta (q_\perp)]  [\varrho D(p_1)],
\end{gather}
To obtain the three-dimensional BSE, we follow Salpeter's methods in Ref.\,\cite{Salpeter1952}. First, we split out $q_P$ from the propagators $S(p_2)$ and $D(p_1)$ as,
\begin{equation}
\begin{gathered}
S(p_2)=-i\left[\frac{\Lambda^+(p_{2\perp})}{q_P-\zeta_2^+-i\epsilon }+\frac{\Lambda^-(p_{2\perp})}{q_P-\zeta_2^-+i\epsilon}\right], \\
\varrho D(p_1)=i \left( \frac{1}{q_P-\zeta_1^++i\epsilon }+\frac{1}{q_P-\zeta_1^--i\epsilon}\right), 
\end{gathered}
\end{equation}
where the poles in propagators are defined as,
\[
\zeta_2^\pm=\alpha_2 M \mp w_2,\quad \zeta_1^\pm=-\alpha_1 M \pm w_1.
\]
Performing the contour integral over $q_P$ on both sides of \eref{E-BS-BD0-1},
%\begin{equation}
% \begin{aligned}
%\varphi (q_\perp)
%&=-i\int \frac{\up{D}q_P}{2\pi}B(q)
%%&= -i\int \frac{\up{D}q_P}{2\pi} \left[\frac{\Lambda^+}{q_P-\zeta_2^+-i\epsilon }+\frac{\Lambda^-}{q_P-\zeta_2^-+i\epsilon}\right]  \Gamma (q_\perp) \left[\frac{1}{q_P-\zeta_1^++i\epsilon }+\frac{1}{q_P-\zeta_1^--i\epsilon}\right]\\
%=\frac{\Lambda^+ \Gamma (q_\perp)}{\zeta^+_2-\zeta^+_1} + \frac{\Lambda^- \Gamma (q_\perp) }{\zeta^-_1-\zeta^-_2},
%%&=\frac{\Lambda_1^+ \vartheta^{ab}\Gamma_b (P,q_\perp)}{M-\w_1-\w_2} - \frac{\Lambda_1^- \vartheta^{ab}\Gamma_b (P,q_\perp) }{M+\w_1+\w_2},
%\end{aligned}
%\end{equation}
we obtain the three-dimensional Salpeter equation of a baryon with $0^+$ diquark core,
 \begin{align}\label{E-BSB-D0}
\varphi (q_\perp)
=\frac{\Lambda^+ \gamma^0  \Theta (q_\perp)}{M-w_1-w_2} - \frac{\Lambda^- \gamma^0 \Theta (q_\perp) }{M+w_1+w_2}.
\end{align}
Using the properties of the projector operators $\Lambda^\pm$, the equation above can be rewritten as two coupled equations,
\begin{equation}
\begin{aligned} 
\varphi^{+}(q_\perp)&\equiv \Lambda^+   \gamma^0  \varphi=+\frac{\Lambda^+ \gamma^0 \Theta (q_\perp)}{M-w_1-w_2},\\
\varphi^{-}(q_\perp)&\equiv \Lambda^-   \gamma^0  \varphi=-\frac{\Lambda^- \gamma^0 \Theta (q_\perp) }{M+w_1+w_2},
\end{aligned}
\end{equation}
where $\varphi =\varphi^{+} +\varphi^{-} $, and $\varphi^+$ and $\varphi^-$ are called the positive and negative energy Salpeter wave functions respectively. These are the coupled baryon Salpeter equations. Simplifying these two coupled equations we obtain the Schr{\"o}dinger-type \eref{E-BSB-D0-2} of the baryons.

\subsection{\textup{Normalization of the Salpeter wave functions for a baryon with $0^+$ diquark}} \label{A-Norm}
The normalization of the BS wave function $B(P,q,\xi)$ is expressed as,
\begin{equation}\label{E-Norm}
-i\int \int \frac{\d^4 q}{(2\pi)^4}\frac{\d^4 k}{(2\pi)^4}  \bar{B}(q,\xi') \frac{\partial}{\partial P^0}  \up{I}(P,q,k) B(k,\xi)  =2M \delta_{\xi\xi'},
\end{equation}
where $\xi(\xi')$ denotes the polarization state of the baryon; the operator $I(P,q,k)$ has the following form,
\begin{equation}
\up{I}(P,q,k)=\up{I}_\up{propagator}+\up{I}_\up{kernel}=S^{-1}(p_2)D^{-1}(p_1)(2\pi)^4 \delta^4(k-q) + iK(p_1,k_1),
\end{equation}
from which we can see that, the normalization in \eref{E-Norm} consists of two parts, the $\up{N_{propagator}}$ part, contributions from the inverses of the propagators, and the $\up{N_{kernel}}$ part, contributions from the interaction kernel. Also notice the dependence of the kernel $iK$ on the $P^0$ is introduced by the factor $\varrho\equiv 2(\alpha_1 M+q_P)$ in the diquark form factor $\mathcal{A}_\mu$, which is different from that in a meson system.

%The BS vertex can be further expressed by the inverse of the propagators as 
%\[
%\Gamma(q)%=-i\int \frac{\up{D}^4 k}{(2\pi)^4} V(P,k-q)g^{\alpha\beta}B_{\beta}(P,k)
%=S^{-1}(p_2)D^{-1}(p_1)B(q).
%\]
%Inserting this result into \eref{E-Norm} to calculate the normalization of the Salpeter wave functions.
%The normalization conditions in \eref{E-Norm-D0} consists of two parts, which is different from the case in the Salpeter equation of the meson.
Using the definition of the instantaneous vertex $\Theta$, the integration $\up{N_{kernel}}$ involved the kernel $iK(p_1,k_1)$ is calculated as
\begin{equation}\notag
\begin{aligned}
% -i\int \int \frac{\up{D}^4 q}{(2\pi)^4}\frac{\up{D}^4 k}{(2\pi)^4} \bar{B}(P,q,\xi') \frac{\partial}{\partial P^0} [iK(p_1,k_1)] B(P,k,\xi) 
%= &  -i2\alpha_1\int \frac{\up{D}^4 q}{(2\pi)^4}   \bar{B}(P,q,\bar r) i \int \frac{\up{D}^4 k}{(2\pi)^4} V(k_\perp-q_\perp) B(P,k,r) \\
%= &  -2\alpha_1\int \frac{\up{D}^3 q_\perp}{(2\pi)^3}   \bar{\varphi}(P,q_\perp,\bar r) \int \frac{\up{D}^3 k_\perp}{(2\pi)^3} V(k_\perp-q_\perp) \varphi(P,k_\perp,r) \\
\up{N_{kernel}}=  -2\alpha_1\int  \frac{\d^3 q_\perp}{(2\pi)^3}  \bar{\varphi}(q_\perp,\xi') \gamma^0 \Theta(q_\perp,\xi),
\end{aligned}
\end{equation}
where we used the result 
$\frac{\partial}{\partial P^0} K(p_1,k_1)= (2\alpha_1)\varkappa(k_\perp - q_\perp)\gamma^0$.
The normalization part involved the propagators' inverses is 
\begin{equation}\notag
\begin{aligned}
%&- i\int \int \frac{\up{D}^4 q}{(2\pi)^4}\frac{\up{D}^4 k}{(2\pi)^4}  \bar{B}(P,q,\xi') \frac{\partial}{\partial P^0} \left[S^{-1}(p_2)D^{-1}(p_1)(2\pi)^4\delta^4(k-q) \right] B(P,k,\xi) \\
%=&-i\int \frac{\up{D}^4 q}{(2\pi)^4}  \bar{B}(P,q,\bar r) \frac{\partial}{\partial P^0} \left[S^{-1}(p_2)D^{-1}(p_1)\right] B(P,q,r) \\
\up{N_{propagator}}=&\int \frac{\d^3 q_\perp}{(2\pi)^3}  \left [ 4\alpha_1 \bar\varphi(q_\perp,\xi') \gamma^0 \Theta(q_\perp,\xi) + 2w_1 \bar\varphi(q_\perp,\xi') \gamma^0 \varphi(q_\perp,\xi) \right].
\end{aligned}
\end{equation}
%where $\varphi_a(s)$ and $\Gamma^{a(c)}(s)$'s dependence on $P$ and $q_\perp$ are skipped for simplicity in the last two lines. 
On the other hand, from \eref{E-BSB-D0-2} the instantaneous vertex $\Theta(q_\perp)$ can also be expressed by the Salpeter wave function as,
\begin{equation} \label{E-BSB-vertex-D0-2}
\Theta (q_\perp) = \left[ M H_2(p_{2\perp}) -(w_1+w_2) \right] \varphi(P,q_\perp).
\end{equation}
Finally, putting the two parts together, we obtain the normalization condition of the baryon's Salpeter wave function as \eref{E-Norm-D0}.

\acknowledgments
We thank Hai-Yang Cheng, Xing-Gang Wu, Xu-Chang Zheng, Tian-Hong Wang and Hui-Feng Fu for the helpful discusses and suggestions. This work is supported by the National Natural Science Foundation of China\,(NSFC) under Grant Nos.\,12005169, 12075301, 11821505, 12047503, 11805024, and 12075073. It is also supported by the Natural Science Basic Research Program of Shaanxi\,(Program No.\,2021JQ-074), and the Fundamental Research Funds for the Central Universities.

%\paragraph{Note added.} This is also a good position for notes added
%after the paper has been written.


%\bibliographystyle{../BST-QIANG}
%%\biboptions{numbers,sort&compress}
%\setlength{\bibsep}{0.8ex}  % vertical spacing between references
%\bibliography{../reference-QIANG}

\end{document}